\documentclass[showpacs,eqsecnum,preprintnumbers]{revtex4}
\usepackage{amsmath}

\begin{document}

\title{Baryon distribution amplitude: Large-$N_{c}$ factorization,
spin-flavor symmetry and soft-pion theorem}
\author{P.V. Pobylitsa}
\affiliation{Institute for Theoretical Physics II, Ruhr University Bochum, D-44780
Bochum, Germany\\
Petersburg Nuclear Physics Institute, Gatchina, St. Petersburg, 188300,
Russia}

\begin{abstract}
The $1/N_{c}$ expansion for the baryon distribution amplitude is constructed
in terms of a specially designed generating functional. At large $N_{c}$
this functional shows exponential behavior. The exponential factor is
universal for all low-lying baryons and baryon-meson scattering states.
Simple factorization properties are established for the preexponential
term. This factorization agrees with the large-$N_{c}$ contracted
$SU(2N_f)$ spin-flavor symmetry.
The consistency of the factorization with the soft-pion theorem
for the baryon distribution amplitude is explicitly checked. A relation
between the generating functionals for the distribution
amplitudes of the nucleon and the $\Delta$ resonance is derived.
\end{abstract}

\maketitle

\section{Introduction}

\subsection{Baryon distribution amplitude in large-$N_{c}$ QCD}

Baryon distribution amplitude $\Psi ^{B}(x_{1},x_{2},x_{3})$ carries
information on the 3-quark component of the baryon in the infinite momentum
frame. The variables $x_{k}$ have the meaning of the momentum fraction of
the baryon carried by the $k$th quark. The baryon distribution amplitude
plays an important role in the QCD analysis of hard exclusive phenomena 
\cite{Lepage-Brodsky-80,CZ-84,BL-89}.

Perturbative QCD\ provides information about the distribution amplitude only
in the asymptotic regime of large scales $\mu $. At finite $\mu$, the
problem of the calculation of the baryon distribution amplitude
belongs to nonperturbative physics. The amount of methods which allow
us to go beyond the bounds of perturbative QCD (without leaving the solid
ground of QCD) is rather small. The limit of the large number of colors 
$N_{c}$ and the $1/N_{c}$ expansion 
\cite{Hooft-74,Hooft-74-B,Coleman-85,Witten-79-80,Witten-79}
belong to this type of method. The problem of the calculation of the
baryon distribution amplitude in the $1/N_{c}$ expansion was raised in 
Refs. \cite{PP-03,DP-04}.

In QCD with $N_{c}$ quark colors, the minimal quark component of the baryon
contains $N_{c}$ quarks so that the baryon distribution amplitude $\Psi
^{B}(x_{1},x_{2},\ldots ,x_{N_{c}})$ depends on $N_{c}$ variables $x_{k}$.
It hardly makes sense to speak about the large-$N_{c}$ asymptotic behavior
of a function depending on $N_{c}$ variables. In order to construct the 
$1/N_{c}$ expansion for the baryon distribution amplitude, one has to
introduce an object which contains the same information as the distribution
amplitude but allows for a systematic $1/N_{c}$ expansion. In 
Ref. \cite{Pobylitsa-2004} it was suggested to use the generating functional
$\Phi _{B}(g)$ depending on an arbitrary ``source'' $g(x)$. In an
oversimplified form its definition is
\begin{equation}
\Phi _{B}(g)\sim \int \Psi ^{B}(x_{1},x_{2},\ldots ,x_{N_{c}})g(x_{1})g(x_{2})\ldots
g(x_{N_{c}})dx_{1}dx_{2}\ldots dx_{N_{c}}\,\,.
\end{equation}
The precise definition of $\Phi _{B}(g)$ requires more care about 
$N_{c}$-dependent normalization factors, spin and flavor indices, etc.
(see Sec.~\ref{BDA-section}).

As was shown in Ref. \cite{Pobylitsa-2004}, the functional $\Phi _{B}(g)$ has an
exponential behavior at large $N_{c}$:
\begin{equation}
\Phi _{B}(g)\overset{N_{c}\rightarrow \infty }{\sim }\exp \left[ N_{c}W(g)
\right] \,.  \label{Phi-B-exponential}
\end{equation}
Functional $W(g)$ has several important properties 
\cite{Pobylitsa-2004}:

1) The same functional $W(g)$ describes all
low-lying baryons as well as baryon-meson scattering states [universality
of $W(g)$].

2) Functional $W(g)$ obeys a nonlinear evolution equation which can be
solved analytically in the asymptotic regime.

3) The problem of the diagonalization of the anomalous dimensions of the
leading-twist baryon operators can be formulated (and solved analytically)
in terms of the nonlinear evolution equation for the functional $W(g)$.

The \emph{exponential} behavior controlled by the functional $W(g)$ is a
peculiar feature of the problem of the baryon distribution amplitude in
large-$N_{c}$ QCD. The traditional methods of the $1/N_{c}$ expansion were
developed for quantities with a \emph{power} large-$N_{c}$ behavior. 
In Ref. \cite{Pobylitsa-2004} the analysis of the exponential
large-$N_{c}$ behavior was based
on the nonlinear evolution equation for the functional $W(g)$.

Although the approach based on the nonlinear evolution equation has allowed one
to establish rather interesting properties of the $1/N_{c}$ expansion, this
method has certain limitations: The evolution equation has a perturbative
origin. One can inject the nonperturbative physics into the evolution
equation as an initial condition at some normalization point but one cannot
extract the nonperturbative physics from the evolution equation.

In this work we use nonperturbative tools for the analysis of the
large-$N_c$ structure of the generating functional $\Phi_B(g)$.
The standard large-$N_{c}$ methods developed for quantities with the
power large-$N_{c}$ behavior include a very important ingredient, so-called
contracted $SU(2N_f)$ spin-flavor symmetry
\cite{Bardakci-84,GS-84,DM-93,Jenkins-93,DJM-94,DJM-95}. In this paper, we will
study manifestations of the spin-flavor symmetry in the $1/N_{c}$
expansion of the functional $\Phi _{B}(g)$.
Another nonperturbative approach to the analysis of the large-$N_c$
limit of $\Phi_B(g)$ is based on the soft-pion theorem.

As was mentioned above, the exponential behavior
(\ref{Phi-B-exponential}) is controlled by the universal functional $W(g)$.
However, we also have the preexponential functional $A_{B}(g)$:
\begin{equation}
\Phi _{B}(g)\sim A_{B}(g)\exp \left[ N_{c}W(g)\right] \,.
\label{Phi-A-naive}
\end{equation}
The functional $A_{B}(g)$ depends on the type of the baryon $B$. The
structure of this dependence is one of the subjects of this paper.

\subsection{Methods and results}

In Ref. \cite{Pobylitsa-2004} the functionals $W(g)$ and $A_{B}(g)$ were
computed in the asymptotic limit of a large normalization scale $\mu $.
However, in the general case of a finite scale $\mu $, the problem of the
calculation of the functionals $W(g)$ and $A_{B}(g)$ belongs to the same
class of difficulty as solving large-$N_{c}$ QCD.

One can introduce the baryon distribution amplitude not only for stable
baryons but also for baryon resonances and for baryon-meson scattering
states (including one baryon and an arbitrary amount of mesons). In some sense
the functional $A_{B}(g)$ can be considered as a sort of
special large-$N_c$ representation for the baryon and baryon-meson states
$|B\rangle$.

The large-$N_{c}$ classification of baryon-meson scattering states
$|B\rangle $ is well known 
\cite{HEHW-84,MK-85,MP-85,Mattis-89,MM-88,MB-89,DPP-88}. 
Historically this problem was first studied in the framework
of the Skyrme model
\cite{Skyrme-61,Witten-83,Adkins-83}
but from the very beginning it was obvious
that the case of large-$N_c$ QCD is similar.
Usually the large-$N_{c}$ analysis is applied to the
scattering states containing one
baryon and one meson. However, this construction can be easily extended to the
case of multi-meson scattering states.
Up to subtleties related to zero modes, one can
represent these scattering states as
\begin{equation}
|B\rangle=\prod\limits_{k}a_{\beta_k}^{+}|B_{0}\rangle\,,
\label{B-state-a}
\end{equation}
where the operators $a_{\beta_k}^{+}$ create scattering states for single mesons
and $|B_{0}\rangle$ is the pure baryon state.

In this paper, we show that the functional $A_{B}(g)$ corresponding to the
state (\ref{B-state-a}) factorizes into a product of the 
``baryon ground state'' functional $A_0(g)$ and
functionals $\Xi_{\beta_k}(g)$ corresponding to the elementary 
$a_{\beta_k}^{+}$ excitations. The
oversimplified form of this factorization is
\begin{equation}
A_{B}(g)=A_{0}(g)\prod\limits_{k}\Xi _{\beta_k}(g)\,.
\label{A-B-naive-factorization}
\end{equation}
However, the real structure is more complicated because of zero modes. The
full description of $O(N_{c}^{0})$ excitations requires the introduction of
extra degrees of freedom in addition to the $a_{\beta }^{+}$ excitations.
For example, in the case of QCD with a $SU(2)$ isospin symmetry, the additional
degrees of freedom are represented by a rotator whose states can be
described by wave functions $\psi (U)$ depending on the $SU(2)$ matrices $U$.
As a result, the baryon and baryon-meson states can be represented at
large $N_{c}$ as linear combinations of the states
\begin{equation}
|B^{\prime }\rangle =\psi (U)\prod_{k}a_{\beta_k}^{+}|B_{0}\rangle \,.
\label{psi-U-a-state}
\end{equation}

The appearance of this $U$ degree of freedom leads to a modification
of the naive factorization (\ref{A-B-naive-factorization}) 
of the functional $A_{B}(g)$.
In this paper, we show that the presence of the factor $\psi(U)$ in
expression (\ref{psi-U-a-state}) for the state $|B^{\prime}\rangle$ has a
very simple manifestation in the corresponding functional $A_{B^{\prime}}(g)$
describing the distribution amplitude of the state $B'$:
\begin{equation}
A_{B^{\prime}}(g)=\psi\left[ Q(g)\right] A_{0}(g)\prod\limits_{k}\Xi_{
\beta_k}(g)\,,  \label{A-B-prime-g}
\end{equation}
Here $Q(g)$ is some $2\times 2$ matrix functional of $g$. This functional is ``universal'' 
in the sense that it does not depend on the state $B'$.
The functional $Q(g)$ plays the same role
for functionals $A_{B^{\prime}}(g)$
in Eq. (\ref{A-B-prime-g}) as the matrix $U$ for the
states (\ref{psi-U-a-state}). However, there is one important difference.
Although matrix $U$ is unitary in Eq.~(\ref{psi-U-a-state}), the matrix
functional $Q(g)$ is not unitary but still belongs to the group
$SL(2,C)$. Therefore $\psi\left[ Q(g)\right] $ should be understood in terms
of the analytical continuation of $\psi(U)$ from $SU(2)$ to $SL(2,C)$. This
analytical continuation makes no problems, since in practical work the role
of functions $\psi(U)$ is played by the $SU(2)$ Wigner $D$ functions, whose
complexification for the $SL(2,C)$ case is straightforward.

We stress that Eqs. (\ref{psi-U-a-state}) and (\ref{A-B-prime-g}) expressing
the correspondence between the states $|B^{\prime }\rangle $ and functionals 
$A_{B^{\prime }}(g)$ are written in an oversimplified form. The precise form
of these equations can be found in Sec.~\ref{A-B-representation-section}.

Now let us turn to the methods which allow us to derive the factorization
property (\ref{A-B-prime-g}). As is well known, large-$N_{c}$ QCD is not
solved. The ``derivations'' of most of the results in large-$N_{c}$ QCD are
based on the extrapolation of the perturbative large-$N_{c}$ counting to the
nonperturbative region, on various models (naive quark model, Skyrme model,
chiral quark soliton model, etc.) and on checks of the consistency with current
algebra, soft-pion theorems, etc.

Our argumentation in favor of the factorization (\ref{A-B-prime-g}) also
cannot be considered as an impeccable final proof. Nevertheless we can
present a rather wide set of arguments. Some of them have been already
suggested in Ref. \cite{Pobylitsa-2004}:

\begin{itemize}
\item Factorization (\ref{A-B-prime-g}) is consistent with the evolution
equation for the functional $A_{B}(g)$.

\item This factorization was observed in the asymptotic regime of large
normalization scales $\mu$, where the functional $A_{B}(g)$ can be computed
analytically.

\item The factorization of the type (\ref{A-B-prime-g}) can be seen in the
naive quark model.
\end{itemize}

In Ref.~\cite{Pobylitsa-05b} factorization (\ref{A-B-prime-g})
was studied in large-$N_c$ models with the following results:
\begin{itemize}
\item
The factorization was checked using the reduction of the large-$N_c$ problem
to the analysis of effective classical dynamical systems
\cite{Berezin-78,Yaffe-82}.
\item The connection was established between this factorization 
and the traditional methods of the $1/N_c$ expansion based on the Hartree
and RPA (random phase approximation) equations \cite{GLM-65,Pobylitsa-03}.
\end{itemize}

In this paper, we describe a new group of arguments:
\begin{itemize}
\item Factorization (\ref{A-B-prime-g}) is consistent with the large-$N_c$
contracted $SU(2N_{F})$ spin-flavor symmetry.

\item This factorization is consistent with the soft-pion theorem for the
distribution amplitude of pion-baryon scattering states \cite{PPS-01}.
\end{itemize}

As is well known, in QCD with the $SU(N_{f})$ flavor symmetry, the
large-$N_{c}$ limit leads to the additional so-called contracted $SU(2N_{f})$
symmetry
\cite{Bardakci-84,GS-84,DM-93,Jenkins-93,DJM-94,DJM-95}.
The usual rotational group  $SO(3)\sim SU(2)$ and the flavor group
$SU(N_{f})$ are described by the standard Lie algebra
\begin{align}
\lbrack J_{k},J_{l}]  &  =i\varepsilon_{klm}J_{m}\,,\label{J-J-commutator}\\
\lbrack T_{a},T_{b}]  &  =if_{abc}T_{c}\,,\label{T-T-commutator}\\
\lbrack J_{k},T_{a}]  &  =0 \label{J-T-commutator}
\end{align}
with the angular momentum generators $J_{k}$ and the flavor generators $T_{a}$.
As usual, summation over repeated indices is implied.

In the large-$N_{c}$ limit, new symmetry generators $X_{ia}$ appear with the
commutation relations
\begin{align}
\lbrack T_{a},X_{jb}] &  =if_{abc}X_{jc}\,,\label{T-X-commutator}\\
\lbrack J_{k},X_{lb}] &  =i\varepsilon_{klm}X_{mb}\,,\label{J-X-commutator}\\
\lbrack X_{ia},X_{jb}] &  =0\,.\label{X-commute}
\end{align}
The Lie algebra (\ref{J-J-commutator})--(\ref{X-commute}) corresponds to the
contracted $SU(2N_{f})$ group. The appearance of this additional symmetry at
large $N_{c}$ is based on general arguments and our large-$N_{c}$ construction
for the generating functional $\Phi_{B}(g)$ must be compatible with this
symmetry. In this paper, we show how the contracted $SU(2N_{f})$ spin-flavor group is
realized in the space of functionals $A_{B}(g)$.

Another consistency check studied in this paper is based on the soft-pion
theorem. Let us consider the state $|B\pi\rangle$ containing baryon $B$ and
pion $\pi$. If pion $\pi$ is soft (in the rest frame of $B$), then the
soft-pion theorem allows us to to reduce the matrix elements of the state
$|B\pi\rangle$ to the matrix elements of the state $|B\rangle$ without the
soft pion. In particular, the generating functional $\Phi_{B\pi}(g)$
describing the distribution amplitude of the state $|B\pi\rangle$ can be
expressed via the generating functional $\Phi_{B}(g)$ of the pure baryon state
$|B\rangle$. Therefore the soft-pion theorem imposes certain constraints on
functionals $A_{B}(g)$ and $A_{B\pi}(g)$. In this paper, we show that these
constraints are compatible with the large-$N_{c}$ factorization properties of
the functionals $A_{B}(g)$.

The structure of the paper is as follows. In Sec.~\ref{BDA-section} we give
a precise definition of the functionals $\Phi _{B}(g)$, $A_{B}(g)$ and
briefly describe the main results of Ref. \cite{Pobylitsa-2004}, including
the universality of $W(g)$ and the evolution equations for functionals $W(g)$
and $A_{B}(g)$. Section~\ref{BM-scattering-states-section} contains the
description of the standard picture of the large-$N_{c}$ baryons and
baryon-meson scattering states. The results considered in 
Sec.~\ref{BM-scattering-states-section} are well known. We concentrate
in Sec.~\ref{BM-scattering-states-section} only on those facts which are needed for our
analysis of the functional $A_{B}(g)$, including the large-$N_{c}$
spin-flavor symmetry
\cite{Bardakci-84,GS-84,DM-93,Jenkins-93,DJM-94,DJM-95}
and the general
properties of the baryon-meson scattering at large $N_{c}$
\cite{HEHW-84,MK-85,MP-85,Mattis-89,MM-88,MB-89,DPP-88}. 
In Sec.~\ref{A-B-representation-section} we turn to the
factorization properties of the functional $A_{B}(g)$ and show that these
properties are completely consistent with the large-$N_{c}$ spin-flavor
symmetry. As a sort of ``practical application'' of the general
factorization equations, we derive an expression for the functional
$A^{(\Delta)}(g)$ corresponding to the $\Delta $ resonance via the nucleon
functional $A^{(N)}(g)$. In Sec.~\ref{SP-theorem-section} we obtain the
large-$N_{c}$ generalization of the soft-pion theorem for the distribution
amplitude of a near threshold pion-nucleon state \cite{PPS-01} and show that
the factorized structure of the functional $A_{B}(g)$ is also consistent
with this soft-pion theorem.

\section{Baryon distribution amplitude at large $N_{c}$}

\label{BDA-section}

\subsection{Definition}

The baryon distribution amplitude $\Psi ^{B}$ is defined as a transition
matrix element of a product of $N_{c}$ light-cone quark fields $\chi
_{f_{k}s_{k}c_{k}}(x_{k})$ taken between the vacuum and the baryon $B$ with
momentum $P$:
\begin{align}
& \Psi _{(f_{1}s_{1})(f_{2}s_{2})\ldots
(f_{N_{c}}s_{N_{c}})}^{B}(x_{1},x_{2},\ldots ,x_{N_{c}})  \notag \\
& =\frac{1}{N_{c}!}\varepsilon _{c_{1}c_{2}\ldots c_{N_{c}}}\langle 0|
\chi_{c_{1}f_{1}s_{1}}(x_{1})
\chi_{c_{2}f_{2}s_{2}}(x_{2})\ldots 
\chi_{c_{N_{c}}f_{N_{c}}s_{N_{c}}}(x_{N_{c}})|B(P)\rangle \,.  \label{BDA-def}
\end{align}
The indices $f_{k},s_{k},c_{k}$ stand for the quark flavor, spin and color,
respectively:
\begin{equation}
f_{k}=1,2,\ldots ,N_{f}\,\,,
\end{equation}
\begin{equation}
s_{k}=\pm \frac{1}{2}\,\,,
\end{equation}
\begin{equation}
c=1,2,\ldots ,N_{c}\,.
\end{equation}
The tensor $\varepsilon _{c_{1}c_{2}\ldots c_{N_{c}}}$ is used for the
antisymmetrization in color. The light-cone quark operators $\chi _{fsc}(x)$
depending on the longitudinal quark momentum fraction $x$ can be expressed
in terms of the usual Dirac quark field $q$ as follows
\begin{equation}
\chi _{cfs}(x)=(nP)^{1/2}\int_{-\infty }^{\infty }\frac{d\lambda }{2\pi }
\bar{u}_{-s}q_{cf}(\lambda n)\exp \left[ i\lambda x(nP)\right] \,.
\label{chi-psi}
\end{equation}
Here $n$ is an auxiliary light-cone vector. The Dirac spinors $u_{s}$
associated with $n$ are normalized by the condition 
\begin{equation}
u_{s}\otimes \bar{u}_{s}=\frac{1}{2}(n\cdot \gamma )(1-2s\gamma _{5})\quad
\left( s=\pm \frac{1}{2}\right) \,.  \label{u-n-density}
\end{equation}
Strictly speaking, expression (\ref{BDA-def}) for the baryon distribution
amplitude is valid only for the light-cone gauge
\begin{equation}
(n\cdot A)=0\,.
\end{equation}
In other gauges, Eq.~(\ref{BDA-def}) must be modified by an insertion of
Wilson lines.

Note that the distribution amplitude defined by Eq.~(\ref{BDA-def}) contains
the momentum conserving delta function:
\begin{equation}
\Psi_{(f_{1}s_{1})(f_{2}s_{2})\ldots(f_{N_{c}}s_{N_{c}})}^{B}(x_{1},x_{2},
\ldots,x_{N_{c}})\sim\delta(x_{1}+x_{2}+\ldots+x_{N_{c}}-1)\,.
\label{Psi-delta-x}
\end{equation}

\subsection{Large-$N_{c}$ limit}

Although the definition (\ref{BDA-def}) of the baryon distribution amplitude 
$\Psi ^{B}$ uses the auxiliary vector $n$ [entering via the operators
$\chi_{cfs}(x)$], the resulting function $\Psi ^{B}$ is independent of $n$.
Usually one normalizes $n$ by the condition $(nP)=1$.
However, at large $N_{c}$ this normalization is not convenient. Indeed, at
$N_{c}\rightarrow \infty $ we have $P=O(N_{c})$ because of the $O(N_c)$ growth
of the baryon mass. If we keep $n$ fixed at
large $N_{c}$, then $(nP)$ will grow as $O(N_{c})$: 
\begin{equation}
n=O(N_{c}^{0})\,,\quad (nP)=O(N_{c})\,.
\end{equation}
At large $N_{c}$ because of the momentum conserving delta function
(\ref{Psi-delta-x}) the distribution amplitude is concentrated at
\begin{equation}
x_{k}\sim 1/N_{c}\,.
\end{equation}
Therefore, it is useful to introduce the new variables 
\begin{equation}
y_{k}=N_{c}x_{k}
\end{equation}
behaving as $O(N_{c}^{0})$. Let us define the function
\begin{equation}
\Psi _{(f_{1}s_{1})(f_{2}s_{2})\ldots (f_{N_{c}}s_{N_{c}})}^{\prime
B}(y_{1},y_{2},\ldots ,y_{N_{c}})=N_{c}^{-N_{c}/2}\Psi^B
_{(f_{1}s_{1})(f_{2}s_{2})\ldots (f_{N_{c}}s_{N_{c}})}\left( \frac{y_{1}}{
N_{c}},\frac{y_{2}}{N_{c}},\ldots ,\frac{y_{N_{c}}}{N_{c}}\right)
\label{Psi-B-prime-Psi-B}
\end{equation}
which is more convenient for the transition to the large-$N_{c}$ limit. The
factor $N_{c}^{-N_{c}/2}$ is inserted here in order to compensate the $N_{c}$
growth of the contribution of the kinematical factor $(nP)^{N_{c}/2}$ coming
from the product of $N_{c}$ fields $\chi $ (\ref{chi-psi}).

Now we define 
\begin{align}
\Phi _{B}(g)& =\sum\limits_{f_{k}s_{k}}\int\limits_{0}^{\infty
}dy_{1}\int\limits_{0}^{\infty }dy_{2}\ldots \int\limits_{0}^{\infty
}dy_{N_{c}}g_{f_{1}s_{1}}(y_{1})g_{f_{2}s_{2}}(y_{2})\ldots
g_{f_{N_{c}}s_{N_{c}}}(y_{N_{c}})  \notag \\
& \times \Psi _{(f_{1}s_{1})(f_{2}s_{2})\ldots (f_{N_{c}}s_{N_{c}})}^{\prime
B}(y_{1},y_{2},\ldots ,y_{N_{c}})\,.  
\label{Phi-def-1}
\end{align}
The distribution amplitude can be expressed via the functional $\Phi _{B}(g)$:
\begin{equation}
\Psi _{(f_{1}s_{1})(f_{2}s_{2})\ldots (f_{N_{c}}s_{N_{c}})}^{B}\left( \frac{
y_{1}}{N_{c}},\frac{y_{2}}{N_{c}},\ldots ,\frac{y_{N_{c}}}{N_{c}}\right) =
\frac{N_{c}^{N_{c}/2}}{N_{c}!}\left( \prod_{k=1}^{N_{c}}\frac{\delta }{
\delta g_{f_{k}s_{k}}(y_{k})}\right) \Phi _{B}(g)\,.  \label{Psi-B-via-Phi-B}
\end{equation}
Therefore, at finite $N_{c}$ the functional $\Phi _{B}(g)$ and the
distribution amplitude $\Psi ^{B}$ contain the same information.

As shown in Ref. \cite{Pobylitsa-2004}, the large-$N_{c}$ asymptotic
behavior of $\Phi_{B}(g)$ is described by the representation
\begin{equation}
\Phi_{B}(g)=N_{c}^{\nu_{B}}A_{B}(g)\exp\left[ N_{c}W(g)\right] \left[
1+O\left( N_{c}^{-1}\right) \right] \,.  
\label{W-def-1}
\end{equation}
Here the naive expression (\ref{Phi-A-naive}) is modified by the factor $N_c^{\nu_B}$.
The properties of this factor are studied in Sec. \ref{nu-B-subsection}.

\subsection{Universality of $W(g)$ and factorization of $A_{B}(g)$}

The functional $W(g)$ appearing in the large-$N_{c}$ decomposition 
(\ref{W-def-1}) is universal for all low-lying baryons [including resonances with 
$O(N_{c}^{-1})$ and $O(N_{c}^{0})$ excitation energies]. The same functional 
$W(g)$ describes the exponential large-$N_{c}$ behavior of the distribution
amplitudes of baryon-meson scattering states containing one baryon and an
arbitrary $O(N_{c}^{0})$ amount of mesons. Moreover, this functional $W(g)$
also describes the large-$N_{c}$ behavior of higher-twist distribution
amplitudes with a finite $O(N_{c}^{0})$ amount of extra insertions of
color-singlet light-cone quark or gluon operators in addition to the product
of $N_{c}$ quark fields on the RHS of Eq.~(\ref{BDA-def}).

The universality of $W(g)$ was discussed in detail in
Ref. \cite{Pobylitsa-2004}. In this paper, we concentrate on the properties of
the preexponential functional $A_{B}(g)$ appearing in the large-$N_{c}$
representation (\ref{W-def-1}). In contrast to $W(g)$, the functional
$A_{B}(g)$ depends on the baryon (or baryon-meson) state $B$. Perturbative
QCD can provide us information on $A_{B}(g)$ only in the asymptotic limit of
the high normalization scale $\mu \rightarrow \infty $. At finite $\mu$, the
functional $A_{B}(g)$ is determined by the nonperturbative dynamics of QCD
and we cannot compute it from the first principles. However, the
large-$N_{c} $ limit allows us to establish certain important properties of
$A_{B}(g)$. It is well known that the baryon resonances with excitation
energies $O(N_{c}^{0})$ and the baryon-meson scattering states can be described
in terms of ``elementary excitations''
\cite{HEHW-84,MK-85,MP-85,Mattis-89,MM-88,MB-89}.
The complete description of the spectrum of the low-lying baryon and
baryon-meson states also includes ``zero modes'' responsible for the
``fine'' $O(N_{c}^{-1})$ structure of the spectrum. In this paper, we show
that the functional $A_{B}(g)$ corresponding to the baryon (baryon-meson)
state $B$ factorizes into a product of functionals associated with ``elementary
excitations'' making the state $B$. The precise form of this factorization
depends on the number of quark flavors $N_{f}$. In fact, because of the presence
of zero modes, the functional $A_{B}(g)$ is a polynomial
of elementary functionals rather than the simple product (\ref{A-B-prime-g}).

\subsection{Evolution equation}

The evolution equations for functionals $W(g)$ and $A_{B}(g)$ were derived
and studied in detail in Ref. \cite{Pobylitsa-2004}. In this section we
present a brief derivation of these equations which are needed for the
analysis of the preexponential factors $A_{B}(g)$.

The dependence of the baryon distribution amplitude 
$\Psi^{B,\mu}(x_{1}\ldots x_{N_{c}})$ on the normalization point
$\mu $ is described by
the well known evolution equation \cite{Lepage-Brodsky-80}
\begin{equation}
\mu \frac{\partial }{\partial \mu }\Psi^{B,\mu }(x_{1},\ldots, x_{N_{c}})=-
\frac{N_{c}+1}{2N_{c}}\frac{\alpha _{s}(\mu )}{\pi }\sum\limits_{1\leq
i<j\leq N_{c}}\bar{K}_{ij}\Psi^{B,\mu }(x_{1},\ldots, x_{N_{c}})\,.
\label{evolution-2}
\end{equation}
The evolution kernel has the simple form of a sum of ``pair interactions'' 
$\bar{K}_{ij}$ between the $i$th and $j$th quark. In this paper, we use
notation $\bar{K}$ for the two-particle evolution kernel (instead of $K$ as
it was in Ref. \cite{Pobylitsa-2004}) in order to avoid confusion with the
eigenvalues of the operator $\mathcal{K}$, which plays an important role in
the spin-flavor symmetry classification of large-$N_{c}$ baryons 
(Sec.~\ref{Operator-K-section}).

This evolution equation can be rewritten in terms of the generating
functional $\Phi_{B}^{\mu}(g)$ (\ref{Phi-def-1})
\begin{equation}
\mu\frac{\partial}{\partial\mu}\Phi_{B}^{\mu}(g)=-\frac{N_{c}+1}{2N_{c}}
\frac{\alpha_{s}(\mu)}{2\pi}\left\{ \left( g\otimes g\right) \cdot\bar {K}
\cdot\left[ \left( \frac{\delta}{\delta g}\otimes\frac{\delta}{\delta g}
\right) \Phi_{B}^{\mu}(g)\right] \right\} \,.  \label{evolution-4}
\end{equation}
Here we use the brief notation
\begin{align}
& \left( g\otimes g\right) \cdot\bar{K}\cdot\left[ \left( \frac{\delta }{
\delta g}\otimes\frac{\delta}{\delta g}\right) \Phi_B^{\mu}(g)\right]  \notag
\\
&
=\int\limits_{0}^{\infty}dy_{1}\int\limits_{0}^{\infty}dy_{2}\sum
\limits_{f_{1}f_{2}s_{1}s_{2}}g_{f_{1}s_{1}}(y_{1})g_{f_{2}s_{2}}(y_{2}) 
\left[ \bar{K}\frac{\delta}{\delta g_{f_{1}s_{1}}(y_{1})}\frac{\delta }{
\delta g_{f_{2}s_{2}}(y_{2})}\Phi_B^{\mu}(g)\right] \,.
\end{align}

Inserting the large-$N_{c}$ decomposition (\ref{W-def-1}) into 
Eq.~(\ref{evolution-4}), we obtain a nonlinear evolution equation for the functional
$W_{\mu }(g)$
\begin{equation}
\mu \frac{\partial }{\partial \mu }W_{\mu }(g)=-a(\mu )\left\{ \left(
g\otimes g\right) \cdot \bar{K}\cdot \left[ \frac{\delta W_{\mu }(g)}{\delta
g}\otimes \frac{\delta W_{\mu }(g)}{\delta g}\right] \right\} \,,
\label{W-evolution-mu}
\end{equation}
where 
\begin{equation}
a(\mu )=\lim_{N_{c}\rightarrow \infty }\frac{\alpha _{s}(\mu )N_{c}}{4\pi }
\,.  \label{a-mu-def}
\end{equation}
It is convenient to introduce the new variable $t$ such that 
\begin{equation}
dt=2a(\mu )\frac{d\mu }{\mu }\,.  \label{dt-d-mu}
\end{equation}
Then 
\begin{equation}
\frac{\partial }{\partial t}W(g,t)=-\frac{1}{2}\left\{ \left( g\otimes
g\right) \cdot \bar{K}\cdot \left[ \frac{\delta W(g,t)}{\delta g}\otimes 
\frac{\delta W(g,t)}{\delta g}\right] \right\} \,.
\label{W-evolution-compact}
\end{equation}
In the next order of the $1/N_{c}$ expansion, we obtain the evolution
equation for $A_{B}(g,t)$
\begin{align}
\frac{\partial }{\partial t}\ln A_{B}(g,t)& =-\left\{ \left( g\otimes
g\right) \cdot \bar{K}\cdot \left[ \frac{\delta W(g,t)}{\delta g}\otimes 
\frac{\delta \ln A_{B}(g,t)}{\delta g}\right] \right\}  \notag \\
& +b(t)\left\{ \left( g\otimes g\right) \cdot \bar{K}\cdot \left[ \frac{
\delta W(g,t)}{\delta g}\otimes \frac{\delta W(g,t)}{\delta g}\right]
\right\}  \notag \\
& -\frac{1}{2}\left\{ \left( g\otimes g\right) \cdot \bar{K}\cdot \left[
\left( \frac{\delta }{\delta g}\otimes \frac{\delta }{\delta g}\right) W(g,t)
\right] \right\} \,,  \label{A-subleading-evolution}
\end{align}
where 
\begin{equation}
b(t)=\frac{1}{2}\lim_{N_{c}\rightarrow \infty }\left\{ N_{c}\left[ 1-\frac{
N_{c}+1}{4\pi }\frac{\alpha _{s}(\mu )}{a(\mu )}\right] \right\} \,.
\end{equation}

\section{Baryon-meson scattering states}

\label{BM-scattering-states-section}

In this section we give a brief description of well known properties
of baryons and baryon-meson scattering states in large-$N_{c}$ QCD. Later,
these properties will be used in the analysis of the factorization of
functionals $A_{B}(g)$.

\subsection{$O(N_{c}^{-1})$ excited baryons}

According to the standard picture of baryons in large-$N_{c}$ QCD with
$N_{f}=2$ flavors, the lowest baryons have equal spin $J$ and isospin
$T$ \cite{Balachandran-83,Witten-83}:
\begin{equation}
T=J=\left\{ 
\begin{array}{ll}
\frac{1}{2},\frac{3}{2},\ldots & \mathrm{for\,odd}\,N_{c} \\ 
0,1,2,\ldots & \mathrm{for\,even}\,N_{c}
\end{array}
\right. \,.  \label{TS-sequence}
\end{equation}
The masses of these baryons have a $1/N_{c}$ suppressed splitting: 
\begin{equation}
M_{T=J}=d_{1}N_{c}+d_{0}+N_{c}^{-1}\left[ d_{-1}^{(1)}+d_{-1}^{(2)}J(J+1)
\right] +O(N_{c}^{-2})\,.  \label{M-T-J}
\end{equation}
These baryons belong to the same representation of the spin-flavor symmetry
group \cite{Bardakci-84,GS-84,DM-93,Jenkins-93,DJM-94}
which becomes asymptotically exact at
large $N_{c}$. As a consequence, the functional $W(g)$ defined by
Eq.~(\ref{W-def-1}) is the same for all low-lying baryons (\ref{TS-sequence}). The
dependence on the type of baryons $B$ appears only in the preexponential
factor $A_{B}(g)$.

One should keep in mind that the excited baryons can be unstable with
respect to decays into lower baryons and mesons. The precise spectrum of
stable excited baryons is determined by the mass thresholds which depend on
the quark masses. In the case of the $O(N_{c}^{-1})$ excited baryon
resonances, their width has an additional $1/N_{c}$ suppression compared to
$O(N_{c}^{-1})$ mass splittings. Therefore in the first orders of the
$1/N_{c} $ expansion we can ignore the instability of these particles. In
particular, in the leading order one can treat $O(N_{c}^{-1})$ excited
baryon resonances as asymptotic states of the $S$ matrix of the baryon-meson
scattering.

\subsection{$O(N_{c}^{0})$ excited baryons and baryon-meson scattering states}

Now we turn to the baryons with the $O(N_{c}^{0})$ excitation energy. In this
case the excitation energy is rather large and the problem of the
instability of baryon resonances becomes more serious. Therefore the theory
must be constructed in terms of baryon-meson scattering states. A general
$O(N_{c}^{0})$ excitation corresponds to a scattering state containing one
baryon and an arbitrary (but finite compared to large $N_{c}$) amount of
mesons. In principle, the general picture of the large-$N_{c}$ limit does
not exclude the existence of stable $O(N_{c}^{0})$ excited baryons, although
the real world seems to ignore this possibility.

Anyway, both baryon-meson scattering states [with the $O(N_{c}^{0})$ meson energy
in the c.m. frame] and ``possible'' stable $O(N_{c}^{0})$ excited baryons
can be described by a common large-$N_{c}$ formalism
\cite{HEHW-84,MK-85,MP-85,Mattis-89,MM-88,MB-89,CK-85}.
The energy of these
states is given by the formula

\begin{equation}
E=d_{1}N_{c}+\left( d_{0}+\sum_{\beta}\Omega_{\beta}n_{\beta}\right)
+O(N_{c}^{-1})\,,  \label{M-harmonic}
\end{equation}
where $n_{\beta}=0,1,2,\ldots$ are integer numbers, and the $\Omega_{\beta}$
are $N_{c}$-independent coefficients.

In the case of stable $O(N_{c}^{0})$ excited baryons, the energy $E$ in 
Eq.~(\ref{M-harmonic}) would correspond to the mass of the excited baryon. 
Eq.~(\ref{M-harmonic}) represents the excitation corrections to the mass as a
discrete set of ``elementary'' excitation energies $\Omega_{\beta}$.

In the case of baryon-meson scattering states containing several mesons, the
energies $\Omega_{\beta}$ should be interpreted as energies of single
mesons. Obviously these single-meson energies $\Omega_{\beta}$ have a
continuous spectrum. For the scattering states with the continuous spectrum of
$\Omega_{\beta}$, only the values $n_{\beta}=0,1$ in Eq.~(\ref{M-harmonic})
have physical sense.

In both cases (stable excited baryons and baryon-meson scattering states),
we can interpret the states (\ref{M-harmonic}) as a result of the action
of creation operators $a_{\beta }^{+}$ on the state $|B_{0}\rangle $
corresponding to the lowest baryon
\begin{equation}
\prod\limits_{\beta }\left( a_{\beta }^{+}\right) ^{n_{\beta }}|B_{0}\rangle
\,.  \label{a-dagger-B-states}
\end{equation}
However, the full picture of the $O(N_{c}^{0})$ excitations is more
complicated. Quantum numbers $n_{\beta }$ appearing in Eq.~(\ref{M-harmonic})
do not fix the excited state completely. The complete description of these
states must contain extra degrees of freedom in addition to the creation
operators $a_{\beta }^{+}$ and additional quantum numbers associated with
the new degrees of freedom. The dependence of the energy (\ref{M-harmonic})
on these extra quantum numbers appears only in the order $O(1/N_{c})$.
Therefore Eq.~(\ref{M-harmonic}) remains valid in the corrected
picture of the $O(N_{c}^{0})$ excitations.

The set of the additional quantum numbers responsible for the $O(1/N_{c})$
``fine structure'' of states depends on the number of quark flavors $N_{f}$. Below, we concentrate on the
case $N_{f}=2$ when the additional quantum system corresponds to an $SU(2)$
rotator. The states of this rotator are described by wave functions $\psi
(U) $ depending on the $SU(2)$ matrix $U$:
\begin{align}
UU^{+}& =1\,,   \\
\det U& =1\,.  \label{det-U-one}
\end{align}
The full Hilbert space $\mathcal{H}$ describing the $O(N_{c}^{0})$
excitations is the tensor product of the space $\mathcal{H}_{U}$ of the
rotator wave functions $\psi (U)$ and of the Fock space $\mathcal{H}_{a}$ of
states (\ref{a-dagger-B-states}) associated with operators $a_{\beta }^{+}$
\begin{equation}
\mathcal{H}=\mathcal{H}_{U}\otimes \mathcal{H}_{a}\,.
\label{H-tensor-product}
\end{equation}
The states of $\mathcal{H}$ can be considered as linear combinations of the
states (\ref{a-dagger-B-states}) with $U$ dependent coefficients:
\begin{equation}
\psi (U)\prod\limits_{\beta }\left( a_{\beta }^{+}\right) ^{n_{\beta
}}|B_{0}\rangle \,.  \label{psi-a-state}
\end{equation}
Operators $a_{\beta }^{+}$ obey the standard commutation relations
\begin{equation}
\lbrack a_{\beta },a_{\beta ^{\prime }}^{+}]=\delta _{\beta \beta ^{\prime
}}\,.  \label{a-a-dagger-commutator}
\end{equation}
In fact, $\beta $ is a multiindex made of several quantum numbers which
contain the full information about the corresponding excitation. In the case
of baryon-meson scattering states, index $\beta $ also includes continuous
parameters. Operators $U$ and $a_{\beta }^{+}$ act in different
spaces of the tensor product (\ref{H-tensor-product}). Therefore these
operators commute:
\begin{equation}
\lbrack a_{\beta },U]=[a_{\beta }^{+},U]=0\,.  \label{a-U-commute-0}
\end{equation}
The energy spectrum (\ref{M-harmonic}) is described by the effective
Hamiltonian
\begin{equation}
H=d_{1}N_{c}+d_{0}+\sum_{\beta }\Omega _{\beta }a_{\beta }^{+}
a_{\beta}+O(N_{c}^{-1})\,.  \label{H-a-a-dagger}
\end{equation}

\subsection{Usual rotational and isospin symmetries}
\label{Symmetries-section}

The effective description of the $O(N_{c}^{0})$ baryon excitations
and baryon-meson scattering states is constructed
for the center of mass frame. In this frame we have the freedom of the
3-dimensional space rotations. This means that in the Hilbert space 
$\mathcal{H}$ (\ref{H-tensor-product}) we must have a representation of the
standard angular momentum algebra (\ref{J-J-commutator}) for
the generators $J_{k}$ corresponding to the total angular momentum of the
baryon-meson state (including the orbital momentum and spin).
We also assume the $SU(2)$ flavor symmetry. Its generators $T_{a}$
obey commutation relations (\ref{T-T-commutator}) and
(\ref{J-T-commutator}) with
\begin{equation}
f_{abc}=\varepsilon_{abc}\,.
\end{equation}

The commutators of the generators $J_{k}$, $T_{a}$ with $U$ are
\begin{align}
\left[ J_{k},U\right] & =\frac{1}{2}U\tau _{k}\,,  \label{J-U-commutator} \\
\left[ T_{a},U\right] & =\frac{1}{2}\tau _{a}^{\mathrm{tr}}U\,.
\label{T-U-commutator}
\end{align}
Here $\tau _{a}$ are the Pauli matrices and {\em tr} stands for the matrix
transposition. In the case of finite transformations, we have
\begin{align}
\exp \left( i\omega _{k}J_{k}\right) U\exp \left( -i\omega _{k}J_{k}\right)
& =U\exp \left( i\omega _{k}\tau _{k}/2\right) \,,  \label{U-transform-J} \\
\exp \left( i\omega _{a}T_{a}\right) U\exp \left( -i\omega _{a}T_{a}\right)
& =\exp \left( i\omega _{a}\tau _{a}^{\mathrm{tr}}/2\right) U\,.
\label{U-transform-T}
\end{align}

Now let us turn to the action of generators $J_{k}$, $T_{a}$ on operators 
$a_{\beta}^{+}$. The commutator algebra 
(\ref{a-a-dagger-commutator}) and the form of the Hamiltonian (\ref{H-a-a-dagger})
do not fix operators $a_{\beta}^{+}$ completely, since the excitation energies
$\Omega_{\beta}$ are degenerate. Using this freedom, one can choose the
operators $a_{\beta}^{+}$ so that they commute with $T_{b}$
\begin{equation}
\left[ T_{b},a_{\beta}^{+}\right] =0\,.  \label{T-a-commute}
\end{equation}
Operators $a_{\beta}^{+}$ transform under the spatial $J_k$ rotations 
according to various irreducible representations of $SU(2)$. Therefore, the
subscript $\beta$ can be written as a multiindex
\begin{equation}
\beta=(m,K,K_{3})\,,
\end{equation}
\begin{equation}
a_{\beta}^{+}=a_{mKK_{3}}^{+}\,.
\label{a-beta-KK}
\end{equation}
We use the special notation $K,K_{3}$ for the angular momentum indices,
and $m$ stands for the remaining quantum numbers.
Under the $J_{k}$ rotations, we have
\begin{equation}
\exp\left( i\omega_{k}J_{k}\right) a_{mKK_{3}}^{+}\exp\left( -i\omega
_{k}J_{k}\right)
=\sum_{K_{3}^{\prime}}a_{mKK_{3}^{\prime}}^{+}D_{K_{3}^{\prime}K_{3}}^{K} 
\left( \exp\left( i\omega_{k}\tau_{k}/2\right) \right) \,.
\label{J-a-dagger-commutator}
\end{equation}
Here $D_{K_{3}^{\prime}K_{3}}^{K}$ is the standard Wigner function
corresponding to the $K$ representation of the group $SU(2)$
(see Appendix~\ref{Wigner-functions-appendix}).

According to Eq.~(\ref{T-a-commute}) operators $a_{mKK_{3}}^{+}$ are invariant under
isospin rotations:
\begin{equation}
\exp\left( i\omega_{b}T_{b}\right) a_{mKK_{3}}^{+}\exp\left( -i\omega
_{b}T_{b}\right) =a_{mKK_{3}}^{+}\,.  \label{T-a-dagger-commutator}
\end{equation}
In terms of the new notation $a_{mKK_{3}}^{+}$ (\ref{a-beta-KK}), relations 
(\ref{a-a-dagger-commutator}) and (\ref{a-U-commute-0}) take the form
\begin{equation}
\left[ a_{mKK_{3}},a_{m^{\prime}K^{\prime}K_{3}^{\prime}}^{+}\right]
=\delta_{mm^{\prime}}\delta_{KK^{\prime}}\delta_{K_{3}K_{3}^{\prime}}\,,
\end{equation}
\begin{equation}
\left[ U,a_{mKK_{3}}^{+}\right] =\left[ U,a_{mKK_{3}}\right] =0\,.
\label{a-U-commute}
\end{equation}
The states (\ref{psi-a-state}) can be rewritten in the form
\begin{equation}
\psi(U)\prod\limits_{j=1}^{n}a_{m^{(j)}K^{(j)}K_{3}^{(j)}}^{+}|B_{0}\rangle
\,.  \label{psi-a-dagger-product}
\end{equation}

\subsection{Large-$N_c$ contracted $SU(2N_f)$ spin-flavor symmetry}

In the previous section, we have demonstrated how the usual rotational symmetry $SO(3)\sim
SU(2)$ and the flavor symmetry $SU(N_{f})$ (for $N_{f}=2$ quark flavors)
are realized in the space of baryon-meson scattering states. Now we want to
show that this $SU(2)_{\mathrm{spin}}\otimes SU(N_{f})_{\mathrm{flavor}}$
symmetry can be extended to the large-$N_{c}$ contracted spin-flavor
$SU(2N_{f})$ symmetry. This extension requires the introduction of
new generators $X_{ia}$.

Let us first introduce the $SU(2)$ matrix
\begin{equation}
V=i\tau _{2}U  \label{V-via-U}\,.
\end{equation}
Next we define
\begin{equation}
X_{ia}=\frac{1}{2}\mathrm{Sp}\left( \tau _{a}V\tau _{i}V^{-1}\right) =\frac{1
}{2}\mathrm{Sp}\left( \tau _{2}\tau _{a}\tau _{2}U\tau _{i}U^{-1}\right) \,.
\label{X-def-via-V}
\end{equation}
Since matrix $V$ is unitary, the $X_{ia}$ is an $SO(3)$ matrix. Therefore
\begin{equation}
X_{ia}X_{ib}=\delta _{ab}\,,  \label{XX-1}
\end{equation}
\begin{equation}
X_{ia}X_{ja}=\delta _{ij}\,,
\end{equation}
\begin{equation}
\varepsilon _{ijk}X_{ia}X_{jb}=\varepsilon _{abc}X_{kc}\,,
\label{eps-XX-0}
\end{equation}
\begin{equation}
\varepsilon _{abc}X_{ia}X_{jb}=\varepsilon _{ijk}X_{kc}\,.
\label{eps-XX}
\end{equation}

Using the definition (\ref{X-def-via-V}) of $X_{ia}$ and the commutators
(\ref{J-U-commutator}) and (\ref{T-U-commutator}), one can easily
check that operator $X_{ia}$ obeys the commutator algebra
(\ref{T-X-commutator})--(\ref{X-commute}).

Thus, our operators $J_i$, $T_a$, $X_{ia}$ obey the full set of the Lie
commutators (\ref{J-J-commutator})--(\ref{X-commute})
of the contracted $SU(2N_f)$ symmetry with $N_f=2$.

\subsection{Operator $\mathcal{K}_{i}$}

\label{Operator-K-section}

Let us define the operators
\begin{equation}
\mathcal{K}_{i}=J_{i}+X_{ia}T_{a}\,,  \label{K-def-via-J-X-T}
\end{equation}
\begin{equation}
\mathcal{K}^{2}=\mathcal{K}_{i}\mathcal{K}_{i}\,.
\label{K-cal-squared}
\end{equation}
Using relations 
(\ref{J-J-commutator})--(\ref{X-commute}), (\ref{XX-1})--(\ref{eps-XX}),
one can easily compute the commutators
\begin{equation}
\left[ J_{k},\mathcal{K}_{l}\right] =i\varepsilon _{klm}\mathcal{K}_{m}\,,
\label{J-K-commutator}
\end{equation}
\begin{equation}
\left[ T_{a},\mathcal{K}_{i}\right] =0\,,  \label{T-K-commutator}
\end{equation}
\begin{equation}
\left[ J_{i},\mathcal{K}^{2}\right] =0\,,  \label{J-K2-commutator}
\end{equation}
\begin{equation}
\left[ T_{a},\mathcal{K}^{2}\right] =0\,,  \label{T-K2-commutator}
\end{equation}
\begin{equation}
\left[ \mathcal{K}_{i},X_{jb}\right] =0\,,  \label{K-X-commute}
\end{equation}
\begin{equation}
\left[ \mathcal{K}_{k},\mathcal{K}_{l}\right] 
=i\varepsilon _{klm}\mathcal{K}_{m}\,.  \label{K-k-commutator}
\end{equation}
We stress that the commutation relations
(\ref{J-K-commutator})--(\ref{K-k-commutator}) follow directly from the algebra 
(\ref{J-J-commutator})--(\ref{X-commute}), (\ref{XX-1})--(\ref{eps-XX})
of the operators $J_{i},T_{a},X_{ia}$ and from the definition
(\ref{K-def-via-J-X-T}) of $\mathcal{K}_{i}$. In 
Sec.~\ref{A-B-representation-section} we will find a representation of the algebra
(\ref{T-X-commutator})--(\ref{X-commute}), (\ref{XX-1})--(\ref{eps-XX})
in the space of functionals $A_{B}(g)$
describing the distribution amplitudes of baryon and baryon-meson states. As
a consequence, the commutation relations (\ref{J-K-commutator})--(\ref{K-k-commutator}) will hold automatically for operators $\mathcal{K}_{i}$
acting in the space of functionals $A_{B}(g)$.

Using Eqs. (\ref{J-U-commutator}), (\ref{T-U-commutator}), (\ref{V-via-U}),
and (\ref{X-def-via-V}), we can also derive
\begin{equation}
\left[ \mathcal{K}_{i},V\right] =0\,,  \label{K-V-commute}
\end{equation}
\begin{equation}
\left[ \mathcal{K}_{i},U\right] =0\,.  \label{K-U-commute}
\end{equation}
The commutators with operators $a_{mKK_{3}}^{+}$ can be computed using
Eqs. (\ref{J-a-dagger-commutator}), (\ref{T-a-dagger-commutator}), and
(\ref{a-U-commute}):
\begin{equation}
\left[ X_{jb},a_{mKK_{3}}^{+}\right] =0\,,  \label{X-a-commutator}
\end{equation}
\begin{equation}
\left[ \mathcal{K}_{i},a_{mKK_{3}}^{+}\right] =\left[ J_{i},a_{mKK_{3}}^{+}
\right] \,,  \label{K-a-J-a-commutator}
\end{equation}
\begin{equation}
\exp \left( i\omega _{n}\mathcal{K}_{n}\right) a_{mKK_{3}}^{+}\exp \left(
-i\omega _{n}\mathcal{K}_{n}\right) 
=\sum_{K_{3}^{\prime
}}a_{mKK_{3}^{\prime }}^{+}D_{K_{3}^{\prime }K_{3}}^{K}\left( \exp \left(
i\omega _{n}\tau _{n}/2\right) \right) \,.  \label{K-a-transformation}
\end{equation}

\subsection{Eigenstates of $J^{2},J_{3},T^{2},T_{3},\mathcal{K}^{2}$}

Relations (\ref{J-T-commutator}), (\ref{J-K2-commutator}), 
and (\ref{T-K2-commutator}) show that we have the following set of \emph{commuting}
operators 
\begin{equation}
J^{2},J_{3},T^{2},T_{3},\mathcal{K}^{2}\,.  \label{JTK-set}
\end{equation}
Here we use the short notation (\ref{K-cal-squared}) and 
\begin{equation}
J^{2}=J_{i}J_{i},\,\quad T^{2}=T_{a}T_{a}\,.
\end{equation}

It is easy to construct the eigenstates of operators (\ref{JTK-set}). Let us
first consider the state 
\begin{equation}
|KK_{3}\{K^{(j)}m^{(j)}\}\rangle =\left[ \sum\limits_{K_{3}^{(1)}K_{3}^{(2)}
\ldots K_{3}^{(n)}}\mathcal{C}_{K^{(1)}K_{3}^{(1)};K^{(2)}K_{3}^{(2)};\ldots
;K^{(n)}K_{3}^{(n)}}^{KK_{3}}\prod
\limits_{j=1}^{n}a_{m^{(j)}K^{(j)}K_{3}^{(j)}}^{+}\right] |B_{0}\rangle \,,
\label{KKm-construction}
\end{equation}
where the coefficient $\mathcal{C}_{K^{(1)}K_{3}^{(1)};K^{(2)}K_{3}^{(2)};
\ldots ;K^{(n)}K_{3}^{(n)}}^{KK_{3}}$ is a ``multiparticle Clebsh-Gordan
coefficient'' adding $n$ angular momenta 
$K^{(1)}K_{3}^{(1)},K^{(2)}K_{3}^{(2)},\ldots ,K^{(n)}K_{3}^{(n)}$ and
producing $KK_{3}$. We omit the quantum numbers fixing the details of this
addition procedure.

The baryon  ``ground-state'' $|B_{0}\rangle $ obeys the condition 
\begin{equation}
J_{i}|B_{0}\rangle =T_{a}|B_{0}\rangle =0\,.  \label{J-T-B0-zero}
\end{equation}
Strictly speaking, the state $|B_{0}\rangle $ with $J=T=0$ exists only for
even $N_{c}$. However, formally we can use $|B_{0}\rangle $ for the
construction of other states also in the case of odd $N_{c}$, keeping in
mind the selection rules for the physically allowed values of $J$ and $T$.

Using Eqs.~(\ref{K-def-via-J-X-T}) and (\ref{J-T-B0-zero}), we find 
\begin{equation}
\mathcal{K}_{i}|B_{0}\rangle =0\,.
\end{equation}
Therefore, we have according to Eq.~(\ref{K-a-transformation}) 
\begin{equation}
\mathcal{K}_{3}|KK_{3}\{K^{(j)}m^{(j)}\}\rangle
=K_{3}|KK_{3}\{K^{(j)}m^{(j)}\}\rangle \,,  \label{K3-K-state}
\end{equation}
\begin{equation}
\mathcal{K}^{2}|KK_{3}\{K^{(j)}m^{(j)}\}\rangle
=K(K+1)|KK_{3}\{K^{(j)}m^{(j)}\}\rangle \,.
\end{equation}
Using Eqs. (\ref{J-a-dagger-commutator}) and (\ref{J-T-B0-zero}), we obtain 
\begin{equation}
J_{3}|KK_{3}\{K^{(j)}m^{(j)}\}\rangle =K_{3}|KK_{3}\{K^{(j)}m^{(j)}\}\rangle
\,,
\end{equation}
\begin{equation}
J^{2}|KK_{3}\{K^{(j)}m^{(j)}\}\rangle
=K(K+1)|KK_{3}\{K^{(j)}m^{(j)}\}\rangle \,.
\end{equation}
Taking into account Eqs. (\ref{T-a-commute}) and (\ref{J-T-B0-zero}), we
find 
\begin{equation}
T_{a}|KK_{3}\{K^{(j)}m^{(j)}\}\rangle =0\,.  \label{T-a-K-state}
\end{equation}
Now we construct the state 
\begin{equation}
|TT_{3}JJ_{3}K\{K^{(j)}m^{(j)}\}\rangle =\sum\limits_{T_{3}^{\prime
}K_{3}}C_{TT_{3}^{\prime }KK_{3}}^{JJ_{3}}\sqrt{2T+1}D_{T_{3}T_{3}^{\prime
}}^{T}(U)|KK_{3}\{K^{(j)}m^{(j)}\}\rangle \,,  \label{TJK-state-short}
\end{equation}
where $C_{TT_{3}^{\prime }KK_{3}}^{JJ_{3}}$ is the standard Clebsh-Gordan
coefficient.

We obtain from Eqs. (\ref{U-transform-J}) and (\ref{U-transform-T}):
\begin{align}
\exp \left( i\omega _{n}J_{n}\right) D_{K_{3}K_{3}^{\prime }}^{K}(U)\exp
\left( -i\omega _{n}J_{n}\right) & =D_{K_{3}K_{3}^{\prime }}^{K}\left( U\exp
\left( i\omega _{n}\tau _{n}/2\right) \right) \,,  \label{J-D-U-transform} \\
\exp \left( i\omega _{a}T_{a}\right) D_{K_{3}K_{3}^{\prime }}^{K}\left(
U\right) \exp \left( -i\omega _{a}T_{a}\right) & =D_{K_{3}K_{3}^{\prime
}}^{K}\left( \exp \left( i\omega _{a}\tau _{a}^{\mathrm{tr}}/2\right)
U\right) \,.  \label{T-D-U-transform}
\end{align}
Using the properties (\ref{K3-K-state})--(\ref{T-a-K-state}) of the state
$|KK_{3}\{K^{(j)}m^{(j)}\}\rangle $ and the transformation rules 
(\ref{J-D-U-transform}), (\ref{T-D-U-transform}), we immediately conclude that
the state (\ref{TJK-state-short}) is a common eigenstate of the commuting
operators (\ref{JTK-set}).

Indeed, the state $|KK_{3}\{K^{(j)}m^{(j)}\}\rangle $ has zero isospin
according to Eq.~(\ref{T-a-K-state}). Therefore, the isospin $T,T_{3}$ of the
state (\ref{TJK-state-short}) is generated by $D_{T_{3}T_{3}^{\prime
}}^{T}(U)$.

Next, the factor $D_{T_{3}T_{3}^{\prime }}^{T}(U)$ has zero $\mathcal{K}_{i}$
momentum according to Eq.~(\ref{K-U-commute}). Therefore, the quantum number
$\mathcal{K}$ of the state (\ref{TJK-state-short}) is inherited from
$|KK_{3}\{K^{(j)}m^{(j)}\}\rangle $.

As for the angular momentum $J,J_{3}$ of the state (\ref{TJK-state-short}),
it comes from the angular momentum $T,T_{3}^{\prime }$ of 
$D_{T_{3}T_{3}^{\prime }}^{T}(U)$ and from the angular momentum $K,K_{3}$ of
the state $|KK_{3}\{K^{(j)}m^{(j)}\}\rangle $. The Clebsh-Gordan coefficient 
$C_{TT_{3}^{\prime }KK_{3}}^{JJ_{3}}$ adds these two contributions producing
the total angular momentum $J,J_{3}$ of the state (\ref{TJK-state-short}).

Thus, under spatial rotations and isospin transformations of the state
(\ref{TJK-state-short}) we have 
\begin{equation}
\exp \left( i\omega _{n}J_{n}\right) |TT_{3}JJ_{3}K\{K^{(j)}m^{(j)}\}\rangle
=\sum\limits_{J_{3}^{\prime }}|TT_{3}JJ_{3}^{\prime
}K\{K^{(j)}m^{(j)}\}\rangle D_{J_{3}^{\prime }J_{3}}^{J}\left( \exp \left(
i\omega _{n}\tau _{n}/2\right) \right) \,,
\end{equation}
\begin{equation}
\exp \left( i\omega _{a}T_{a}\right) |TT_{3}JJ_{3}K\{K^{(j)}m^{(j)}\}\rangle
=\sum\limits_{T_{3}^{\prime }}|TT_{3}^{\prime
}JJ_{3}K\{K^{(j)}m^{(j)}\}\rangle D_{T_{3}^{\prime }T_{3}}^{T}\left( \exp
\left( i\omega _{a}\tau _{a}/2\right) \right) \,.
\end{equation}

Inserting Eq.~(\ref{KKm-construction}) into Eq.~(\ref{TJK-state-short}), we
obtain 
\begin{gather}
|TT_{3}JJ_{3}K\{K^{(j)}m^{(j)}\}\rangle =\sum\limits_{T_{3}^{\prime
}K_{3}}C_{TT_{3}^{\prime }KK_{3}}^{JJ_{3}}\sqrt{2T+1}D_{T_{3}T_{3}^{\prime
}}^{T}(U)  \notag \\
\times \left[ \sum\limits_{K_{3}^{(1)}K_{3}^{(2)}\ldots K_{3}^{(n)}}\mathcal{
C}_{K^{(1)}K_{3}^{(1)};K^{(2)}K_{3}^{(2)};\ldots
;K^{(n)}K_{3}^{(n)}}^{KK_{3}}\prod
\limits_{j=1}^{n}a_{m^{(j)}K^{(j)}K_{3}^{(j)}}^{+}\right] |B_{0}\rangle \,.
\label{U-a-state-general-1}
\end{gather}
The excitation energies $\Omega _{\beta }$ (\ref{M-harmonic}) associated
with operators $a_{\beta }^{+}=a_{mKK_{3}}^{+}$ (\ref{a-beta-KK}) depend
only on $m$ and $K$ but not on $K_{3}$. Therefore, the energy of the state
(\ref{U-a-state-general-1}) is

\begin{equation}
E=N_{c}d_{1}+d_{0}+\sum\limits_{j=1}^{n}\Omega_{m^{(j)}K^{(j)}}+O(1/N_{c})\,.
\label{Delta-E-sum-omega}
\end{equation}

Using the standard expression for the Clebsh-Gordan coefficients via the $3j$
symbols [see Eq. (\ref{C-3j}) in Appendix~\ref{Wigner-functions-appendix}],
we can rewrite Eq.~(\ref{U-a-state-general-1}) in the form 
\begin{gather}
|TT_{3}JJ_{3}K\{K^{(j)}m^{(j)}\}\rangle =\sqrt{(2T+1)(2J+1)}
\sum\limits_{T_{3}^{\prime }K_{3}}D_{T_{3}T_{3}^{\prime
}}^{T}(U)(-1)^{T-K+J_{3}}  \notag \\
\times \left( 
\begin{array}{ccc}
T & K & J \\ 
T_{3}^{\prime } & K_{3} & -J_{3}
\end{array}
\right) \left[ \sum\limits_{K_{3}^{(1)}K_{3}^{(2)}\ldots K_{3}^{(n)}}
\mathcal{C}_{K^{(1)}K_{3}^{(1)};K^{(2)}K_{3}^{(2)};\ldots
;K^{(n)}K_{3}^{(n)}}^{KK_{3}}\prod
\limits_{j=1}^{n}a_{m^{(j)}K^{(j)}K_{3}^{(j)}}^{+}\right] |B_{0}\rangle \,.
\label{TJ-state-3j-1}
\end{gather}

\subsection{Simplest states}

The lowest baryon states correspond to $n=0$ in Eq.~(\ref{Delta-E-sum-omega}),
i.e. to the absence of the $a_{m^{(j)}K^{(j)}K_{3}^{(j)}}^{+}$
excitations in Eq.~(\ref{U-a-state-general-1}). In this case we have 
$K=K_{3}=0$ in Eq.~(\ref{U-a-state-general-1}) and the Clebsh-Gordan
coefficient $C_{TT_{3}^{\prime }KK_{3}}^{JJ_{3}}$ leads to the constraint 
\begin{equation}
J=T\,.
\end{equation}
As a result, one arrives at the states 
\begin{equation}
\left| T=J,T_{3}J_{3}\right\rangle =\sqrt{2T+1}D_{T_{3}J_{3}}^{T=J}(U)\,.
\label{lowest-B-states}
\end{equation}
The masses of these states are given by Eq.~(\ref{M-T-J}).

In the case of single $a_{m^{(j)}K^{(j)}K_{3}^{(j)}}^{+}$ excitations
corresponding to $n=1$ in Eq.~(\ref{U-a-state-general-1}), we arrive at the
states of the form 
\begin{equation}
|TT_{3}JJ_{3}Km\rangle =\sum\limits_{T_{3}^{\prime }K_{3}}C_{TT_{3}^{\prime
}KK_{3}}^{JJ_{3}}\sqrt{2T+1}D_{T_{3}T_{3}^{\prime
}}^{T}(U)a_{mKK_{3}}^{+}|B_{0}\rangle \,.
\end{equation}

\section{Factorization properties of functionals $A_{B}(g)$}

\label{A-B-representation-section}

\subsection{Functionals $A_{B}(g)$ as a representation for baryon-meson
states}

\label{Functional-AB-representation-subsection}

In contrast to the universal functional $W(g)$, the functional $A_{B}(g)$
appearing in Eq.~(\ref{W-def-1}) depends on the baryon-meson state 
$|B\rangle $. In some sense, the functionals $A_{B}(g)$ can be considered as a
special representation for the baryon-meson states $|B\rangle $. Strictly
speaking, this point of view is not quite correct because of at least two
reasons.

First, we cannot guarantee that any different baryon-meson states 
$|B_{1}\rangle$ and $|B_{2}\rangle$ generate different functionals 
$A_{B_{1}}(g)$ and $A_{B_{2}}(g)$. Second, the definition (\ref{W-def-1}) of
the functional $A_{B}(g)$ assumes the separation of the power factor 
$N_{c}^{\nu_{B}}$. Generally speaking, the power $\nu_{B}$ depends on the
state $|B\rangle$.

Nevertheless the interpretation of the functionals $A_{B}(g)$ in terms of a
new representation for baryon-meson states is rather helpful for the
understanding of the formal relations which are discussed below. We will
often use this interpretation below (keeping in mind its limitations).

The algebra of operators $a_{mKK_{3}}\,$, $a_{mKK_{3}}^{+}$, and $U$
described in the previous section must have some representation in the space
of functionals $A_{B}(g)$. According to Eq.~(\ref{a-U-commute}), operators $U$
and $a_{mKK_{3}}^{+}$ commute. This means that there must exist a
representation in which $U$ and $a_{mKK_{3}}^{+}$ are diagonal. Our
conjecture is that the representation for baryon-meson states $|B\rangle$ in
terms of functionals $A_{B}(g)$ gives this kind of representation: 
\begin{equation}
U_{fs}A_{B}(g)=Q_{fs}(g)A_{B}(g)\,,  \label{U-g-representation}
\end{equation}
\begin{equation}
a_{mKK_{3}}^{+}A_{B}(g)=\xi_{mKK_{3}}(g)A_{B}(g)\,,  \label{a-dagger-g}
\end{equation}
where $Q_{fs}(g)$ and $\xi_{mKK_{3}}(g)$ are some functionals.

\subsection{Properties of functional $Q_{fs}(g)$}

\label{Q-properties-section}

According to Eq.~(\ref{det-U-one}), the matrix $U$ has a unit determinant.
Therefore relation (\ref{U-g-representation}) immediately leads us to the
conclusion that 
\begin{equation}
\det_{fs}Q_{fs}(g)=1\,.  \label{det-Q-1}
\end{equation}
Although the matrix $Q_{fs}(g)$ belongs to $SL(2,C)$, it is not unitary.
Indeed, the functionals $A_{B}(g)$ are analytical in $g$. Representation
(\ref{U-g-representation}), (\ref{a-dagger-g}) must be compatible with the
analyticity of $A_{B}(g)$. This means that functionals $Q_{fs}(g)$ and $\xi
_{mKK_{3}}(g)$ are also analytical in $g$. The analyticity of $Q_{fs}(g)$
excludes its unitarity.

Below we will often work with Wigner functions $D_{mm^{\prime }}^{j}(R)$
extended to the $SL(2,C)$ matrices $R$. As is well known, the
finite-dimensional irreducible representations of $SL(2,C)$ are parametrized
by two spins $(J_{1},\dot{J}_{2})$. Let us concentrate on the
representations with $\dot{J}_{2}=0$. One can consider these representations
as an analytical continuation of the standard Wigner functions 
$D_{mm^{\prime }}^{J_{1}}(R)$ from $SU(2)$ to $SL(2,C)$. Below we use
notation $D_{mm^{\prime }}^{J_{1}}(R)$ for $SL(2,C)$ matrices $R$ assuming
that $\dot{J}_{2}=0$. It is also well known that Wigner functions 
$D_{mm^{\prime }}^{J}(R)$ can be represented as polynomials in matrix
elements of $R$ if $R$ belongs to $SL(2,C)$. The inverse is also true: Any
polynomial $P(R)$ of matrix elements of $R$ can be decomposed into a linear
combination of functions $D_{mm^{\prime }}^{J}(R)$: 
\begin{equation}
P(R)=\sum\limits_{Jmm^{\prime }}c_{Jmm^{\prime }}D_{mm^{\prime }}^{J}(R)\,.
\end{equation}

Since $Q(g)$ belongs to $SL(2,C)$, we can take $R=Q(g)$ in this
decomposition: 
\begin{equation}
P\left( Q(g)\right) =\sum\limits_{Jmm^{\prime}}c_{Jmm^{\prime}}D_{mm^{
\prime}}^{J}\left( Q(g)\right) \,.  \label{P-Q-D}
\end{equation}
We see that functions $D_{mm^{\prime}}^{J}\left( Q(g)\right) $ make a basis
in the space of polynomials $P\left( Q(g)\right) $.

\subsection{Functional $\Xi_{B}(g)$}

In the large-$N_{c}$ world with even $N_{c}$ we can take the ground state
baryon $B_{0}$ (\ref{J-T-B0-zero}) with $J=T=0$ and denote its functional 
$A_{0}(g,t)$. Now let us define 
\begin{equation}
\Xi _{B}(g)=\frac{A_{B}(g)}{A_{0}(g)}\,.  \label{Xi-AA-def}
\end{equation}
This definition can be used for any $N_{c}$ (including odd values), although
functional $A_{0}(g)$ is taken from the even-$N_{c}$ case. The functional 
$A_{0}(g)$ corresponds to the state $|B_{0}\rangle $ appearing in 
Eq.~(\ref{TJ-state-3j-1}). Note that the baryon state with $J=T=1/2$ is possible only
for odd $N_{c}$ whereas the functional $A_{0}(g)$, used in the definition of 
$\Xi _{B}(g)$ (\ref{Xi-AA-def}), was defined in the large-$N_{c}$ world with
even $N_{c}$. This mixed odd-even $N_{c}$ construction appears because $Q(g)$
describes elementary excitations of the \emph{fermion} type.

If we accept relations (\ref{U-g-representation}) and (\ref{a-dagger-g}),
then the calculation of functionals $\Xi
_{TT_{3}JJ_{3}K\{K^{(j)}m^{(j)}\}}(g)$ for the $O(N_{c}^{-1})$ and 
$O(N_{c}^{0})$ excited baryon and baryon-meson states (\ref{TJ-state-3j-1})
becomes trivial: the state (\ref{TJ-state-3j-1}) is mapped to the
functional
\begin{gather}
\Xi _{TT_{3}JJ_{3}K\{K^{(j)}m^{(j)}\}}(g)=\sqrt{(2T+1)(2J+1)}
\sum\limits_{T_{3}^{\prime }K_{3}}D_{T_{3}T_{3}^{\prime }}^{T}\left(
Q(g)\right) (-1)^{T-K+J_{3}}  \notag \\
\times \left( 
\begin{array}{ccc}
T & K & J \\ 
T_{3}^{\prime } & K_{3} & -J_{3}
\end{array}
\right) \sum\limits_{K_{3}^{(1)}K_{3}^{(2)}\ldots K_{3}^{(n)}}\mathcal{C}
_{K^{(1)}K_{3}^{(1)};K^{(2)}K_{3}^{(2)};\ldots
;K^{(n)}K_{3}^{(n)}}^{KK_{3}}\prod\limits_{j=1}^{n}\xi
_{m^{(j)}K^{(j)}K_{3}^{(j)}}(g)\,.  \label{Xi-Q-xi-1}
\end{gather}
The Wigner function $D_{T_{3}T_{3}^{\prime }}^{T}\left( Q(g)\right) $ should
be understood in the sense of the $SL(2,C)$ complexification of the $SU(2)$
representations, as was explained in Sec.~\ref{Q-properties-section}.

In the case of single excitations ($n=1$) we find from Eq.~(\ref{Xi-Q-xi-1}) 
\begin{gather}
\Xi_{TT_{3}JJ_{3}Km}(g)=\sqrt{(2T+1)(2J+1)}\sum\limits_{T_{3}^{
\prime}K_{3}}D_{T_{3}T_{3}^{\prime}}^{T}\left( Q(g)\right) (-1)^{T-K+J_{3}} 
\notag \\
\times\left( 
\begin{array}{ccc}
T & K & J \\ 
T_{3}^{\prime} & K_{3} & -J_{3}
\end{array}
\right) \xi_{mKK_{3}}(g)\,.  \label{Xi-Q-xi-single}
\end{gather}

The lowest $O(N_{c}^{-1})$ excitations (\ref{lowest-B-states}) correspond to 
$n=0$ in Eq.~(\ref{Xi-Q-xi-1}): 
\begin{equation}
\Xi_{T=J,T_{3}J_{3}}(g)=\sqrt{2T+1}D_{T_{3}J_{3}}^{T}\left( Q(g)\right) \,.
\label{Xi-J-eq-T}
\end{equation}

\subsection{Consistency with the evolution equation}

Let us check the consistency of the representation (\ref{Xi-Q-xi-1}) with
the evolution equation (\ref{A-subleading-evolution}). Indeed, taking the
difference of equations (\ref{A-subleading-evolution}) for an excited state
$A_{B}$ and for $A_{0}$, we find 
\begin{equation}
\frac{\partial \Xi _{B}(g,t)}{\partial t}=-\left\{ \left( g\otimes g\right)
\cdot \bar{K}\cdot \left[ \frac{\delta W(g,t)}{\delta g}\otimes \frac{\delta
\Xi _{B}(g,t)}{\delta g}\right] \right\} \,.  \label{Xi-n-evolution}
\end{equation}
Functionals $\Xi _{B}(g,t)$ (\ref{Xi-AA-def}) are associated with baryon
(baryon-meson) states $|B\rangle $. As explained in
Sec.~\ref{Functional-AB-representation-subsection}, one can think about functionals
$\Xi _{B}(g,t)$ as a special representation for $O(N_{c}^{-1})$ and
$O(N_{c}^{0})$ excited states in large-$N_{c}$ QCD. In particular, we have an
implementation of QCD symmetries in the space of functionals $\Xi _{B}(g,t)$.

Solutions of Eq.~(\ref{Xi-n-evolution}) have an important property. If
$\{\Xi _{n}(g,t)\}$ is some set of solutions, then any function of these
solutions 
\begin{equation}
F(\Xi _{1}(g,t),\Xi _{2}(g,t),\ldots )
\end{equation}
is also a solution of the same equation:
\begin{equation}
\frac{\partial F}{\partial t}=-\left[ \left( g\otimes g\right) \cdot \bar{K}
\cdot \left( \frac{\delta W}{\delta g}\otimes \frac{\delta F}{\delta g}
\right) \right] \,.  \label{F-evolution}
\end{equation}

This property of the solutions of the evolution equation
(\ref{Xi-n-evolution}) leads us to the idea that there must exist a set of
``elementary'' functionals $\Xi _{n}(g,t)$ such that an arbitrary solution
$\Xi _{B}(g,t)$ corresponding to some baryon-meson state can be represented
as a functional of the elementary solutions $\Xi _{n}(g,t)$. Note that the
decomposition (\ref{Xi-Q-xi-1}) gives exactly this type of representation.
Indeed, according to Eq.~(\ref{Xi-Q-xi-1}) the functional $\Xi
_{TT_{3}JJ_{3}K\{K^{(j)}m^{(j)}\}}(g)$ is a linear combination of
functionals 
\begin{equation}
D_{T_{3}T_{3}^{\prime }}^{T}\left[ Q(g,t)\right] \prod\limits_{j=1}^{n}\xi
_{m^{(j)}K^{(j)}K_{3}^{(j)}}(g,t)\,.  \label{D-xi}
\end{equation}
Here we have explicitly marked the dependence of the functionals $Q(g,t)$
and $\xi _{mKK_{3}}(g,t)$ on the scale $t$. Note that $D_{T_{3}T_{3}^{\prime
}}^{T}\left[ Q(g,t)\right] $ can be represented as a polynomial in matrix
elements of $Q(g,t)$. Therefore $\Xi _{TT_{3}JJ_{3}K\{K^{(j)}m^{(j)}\}}(g)$
(\ref{Xi-Q-xi-1}) is a polynomial in ``elementary'' functionals $Q(g,t)$ and
$\xi _{mKK_{3}}(g,t)$. If these elementary functionals obey Eq.
(\ref{Xi-n-evolution}), then the ``composite'' functional $\Xi
_{TT_{3}JJ_{3}K\{K^{(j)}m^{(j)}\}}(g)$ automatically satisfies the same
Eq. (\ref{Xi-n-evolution}) according to the property (\ref{F-evolution}). The
same argument shows that Eq. (\ref{F-evolution}) also holds for
$F=\det Q(g,t)$. This means that the evolution equation is consistent with the
constraint (\ref{det-Q-1}).

Let us summarize. In this section we have explicitly checked the consistency
of Eqs.~(\ref{Xi-n-evolution}) and (\ref{Xi-Q-xi-1}). The evolution
equation (\ref{Xi-n-evolution}) comes from perturbative QCD, whereas 
Eq.~(\ref{Xi-Q-xi-1}) describes the non-perturbative properties of baryons
based on the contracted spin-flavor group. Thus the two different aspects of
hadronic physics perfectly match in our large-$N_{c}$ equations.

\subsection{Symmetries}

\subsubsection{Transformations of $g$}

In Sec.~\ref{Functional-AB-representation-subsection}
we described the action of the
operators $U$ and $a_{mKK_{3}}^{+}$ on functionals $A_{B}(g)$. Now we want
to study the action of operators $J_{i}$ and $T_{a}$. The realization of
the isospin symmetry in the space of functionals $A_{B}(g)$ is
straightforward. Starting from the relation
\begin{equation}
\left[ T_{a},g_{fs}\right] =\frac{1}{2}
(\tau_a)_{f^{\prime }f}g_{f^{\prime }s}\,,
\end{equation}
we obtain
\begin{equation}
\exp \left( i\omega _{a}T_{a}\right) g\exp \left( -i\omega _{a}T_{a}\right)
=\exp \left( i\omega _{a}\tau _{a}^{\mathrm{tr}}/2\right) g
\end{equation}
so that for any functional $F(g)$ we have
\begin{equation}
\left[ \exp \left( i\omega _{a}T_{a}\right) F\exp \left( -i\omega
_{a}T_{a}\right) \right] (g)
=F\left( \exp \left( i\omega _{a}\tau _{a}^{\mathrm{tr}}/2\right) g\right) \,.
  \label{exp-T-F-action}
\end{equation}
With the generators of space rotations $J_{i}$ the situation is more subtle.
In the light-cone formalism the rotational symmetry is ``broken'' by the
light-cone vector $n$ appearing in the definition of the baryon distribution
amplitude (\ref{BDA-def}), (\ref{chi-psi}). Actually this symmetry is not
broken but simply becomes ``hidden''. Therefore, simple transformation rules
in terms of $g$ can be written only for the axial rotations around the vector $n$.
Assuming that this light-cone vector is directed along the third axis, we
can write the transformation rule for axial rotations, which can be applied to
an arbitrary functional $F(g)$:
\begin{equation}
\left[ \exp\left( i\omega_{3} J_{3}\right) F\exp\left( -i\omega_{3} J_{3}\right) 
\right] (g)
=F\left( g \exp \left( i\omega_{3}\tau _{3}/2\right)\right) \,.
\end{equation}

\subsubsection{Transformation rules for $Q(g)$}

Although we cannot write a general formula for the action of generators
$J_{1},J_{2}$ on an arbitrary functional $F(g)$, we know how these generators
act on some special functionals. For example, comparing 
Eqs. (\ref{U-transform-J}) and (\ref{U-transform-T}) with Eq.~(\ref{U-g-representation}),
we conclude that
\begin{align}
\exp\left( i\omega_{k}J_{k}\right) Q(g)\exp\left( -i\omega_{k}J_{k}\right) &
=Q(g)\exp\left( i\omega_{k}\tau_{k}/2\right) \,,  \label{exp-J-Q} \\
\exp\left( i\omega_{a}T_{a}\right) Q(g)\exp\left( -i\omega_{a}T_{a}\right) &
=\exp\left( i\omega_{a}\tau_{a}^{\mathrm{tr}}/2\right) Q(g)\,.
\label{exp-T-Q}
\end{align}
In the infinitesimal form we have
\begin{equation}
\left[ J_{k},Q_{fs}(g)\right] =\frac{1}{2}Q_{fs^{
\prime}}(g)\left( \tau_{k}\right) _{s^{\prime}s}\,,  \label{J-Q-commutator-1}
\end{equation}
\begin{equation}
\left[ T_{a},Q_{fs}(g)\right] =\frac{1}{2}Q_{f^{
\prime}s}(g)\left( \tau_{a}\right) _{f^{\prime}f}\,.
\label{T-Q-commutator-1}
\end{equation}

Combining Eqs. (\ref{exp-T-F-action}) and (\ref{exp-T-Q}), we find
\begin{equation}
Q\left( \exp\left( i\omega_{k}\tau_{k}/2\right) g\right) =\exp\left(
i\omega_{k}\tau_{k}/2\right) Q(g)\,\,.
\end{equation}

\subsubsection{Symmetries of $\protect\xi$}

According to Eqs. (\ref{J-a-dagger-commutator}), (\ref{T-a-dagger-commutator}),
and (\ref{a-dagger-g}) we have
\begin{equation}
\left[ T_{a},\xi_{mKK_{3}}(g)\right] =0\,,  \label{T-xi-convention}
\end{equation}
\begin{equation}
\exp\left( i\omega_{j}J_{j}\right) \xi_{mKK_{3}}(g)\exp\left( -i\omega
_{j}J_{j}\right)
=\sum_{K_{3}^{\prime}}\xi_{mKK_{3}^{\prime}}D_{K_{3}^{\prime}K_{3}}^{(K)} 
\left( \exp\left( i\omega_{j}\tau_{j}/2\right) \right) \,\,,  \label{J-xi}
\end{equation}
Both $\xi_{mKK_{3}}(g)$ and $Q(g)$ are ``multiplication operators''.
Therefore they commute
\begin{equation}
\left[ \xi_{mKK_{3}}(g),Q_{fs}(g)\right] =0\,.  \label{xi-Q-commute}
\end{equation}

\subsection{Spin-flavor symmetry}

By analogy with Eq.~(\ref{V-via-U}), we define
\begin{equation}
\tilde{Q}(g)=i\tau_{2}Q(g)\,.  \label{Q-tilde-def}
\end{equation}
Then we find from Eqs. (\ref{J-Q-commutator-1}) and (\ref{T-Q-commutator-1})
\begin{equation}
\left[ J_{i},\tilde{Q}_{fs}(g)\right] =\frac{1}{2}
\tilde{Q}_{fs^{\prime}}(g)\left( \tau_{i}\right) _{s^{\prime}s}\,,
\label{J-Q-tilde-commutator}
\end{equation}
\begin{equation}
\left[ T_{a},\tilde{Q}_{fs}(g)\right] =-\frac{1}{2}
\left( \tau_{a}\right) _{ff^{\prime}}\tilde{Q}_{f^{\prime}s}(g)\,.
\label{T-Q-tilde-commutator}
\end{equation}

For $SL(2,C)$ matrices $R$, we have the general relation
\begin{equation}
\left( i\tau_{2}R\right) ^{-1}=\left( Ri\tau_{2}\right) ^{\mathrm{tr}}\,.
\end{equation}
Applying this identity to $U$ and $\tilde{Q}(g)$, we obtain
\begin{equation}
\left( i\tau_{2}U\right) ^{-1}=\left( Ui\tau_{2}\right) ^{\mathrm{tr}
}\,,\quad\left[ i\tau_{2}Q(g)\right] ^{-1}=\left[ Q(g)i\tau_{2}\right] ^{
\mathrm{tr}}\,.
\end{equation}
Combining these relations with Eq.~(\ref{U-g-representation}), we find
\begin{equation}
\left[ \left( i\tau_{2}U\right) ^{-1}\right] _{fs}A_{B}(g)=\left\{ \left[
i\tau_{2}Q(g)\right] ^{-1}\right\} _{fs}A_{B}(g)\,.
\end{equation}
In terms of matrices $V$ (\ref{V-via-U}) and $\tilde{Q}(g)$ (\ref{Q-tilde-def}), this
relation becomes
\begin{equation}
\left( V^{-1}\right) _{fs}A_{B}(g)=\left\{ \left[ \tilde{Q}(g)\right]
^{-1}\right\} _{fs}A_{B}(g)\,.  \label{V-inv-Q-tilde-inv}
\end{equation}
Now we derive from Eqs. (\ref{X-def-via-V}), (\ref{U-g-representation}), and
(\ref{V-inv-Q-tilde-inv})
\begin{equation}
X_{ia}A_{B}(g)=\frac{1}{2}\mathrm{Sp}\left\{ \tilde{Q}(g)\tau_{i}\left[ 
\tilde{Q}(g)\right] ^{-1}\tau_{a}\right\} A_{B}(g)\,.
\end{equation}
Thus the action of the operator $X_{ia}$ on $A_{B}(g)$ reduces to the factor
\begin{equation}
X_{ia}=\frac{1}{2}\mathrm{Sp}\left\{ \tilde{Q}(g)\tau_{i}\left[ \tilde {Q}(g)
\right] ^{-1}\tau_{a}\right\} \,.
\end{equation}
Using this expression for $X_{ia}$ and commutation relations
(\ref{J-Q-tilde-commutator}), (\ref{T-Q-tilde-commutator}), it is straightforward
to show that this operator obeys all spin-flavor algebra relations 
(\ref{T-X-commutator})--(\ref{X-commute}), (\ref{XX-1})--(\ref{eps-XX}).
Note
that for the derivation of these relations one does not need the unitarity of
$\tilde{Q}(g)$; it is sufficient that $\tilde{Q}(g)$ belongs to $SL(2,C)$.

Thus, we have shown that our conjecture about the realization 
(\ref{U-g-representation}), (\ref{a-dagger-g}) of the operators $U$ and 
$a_{mKK_{3}}^{+}$ in the space of functionals $A_{B}(g)$ is consistent with
the algebra of the spin-flavor symmetry
(\ref{T-X-commutator})--(\ref{X-commute}), (\ref{XX-1})--(\ref{eps-XX}).

Now, when we have the set of operators $T_{a}$, $J_{i}$, $X_{ia}$ acting on
functionals $A_{B}(g)$ and obeying the algebraic relations
(\ref{J-J-commutator})--(\ref{X-commute}), (\ref{XX-1})--(\ref{eps-XX}),
we can use Eq.~(\ref{K-def-via-J-X-T}) in order to define
operator $\mathcal{K}_{i}$ acting on functionals $A_{B}(g)$. This operator 
$\mathcal{K}_{i}$ automatically obeys the commutation relations 
(\ref{J-K-commutator})--(\ref{K-k-commutator}), since these commutators follow
directly from the algebraic relations
(\ref{J-J-commutator})--(\ref{X-commute}), (\ref{XX-1})--(\ref{eps-XX}),
which have been already checked for our representation of the spin-flavor
algebra in the space of functionals $A_{B}(g)$.

Using relations (\ref{T-xi-convention})--(\ref{xi-Q-commute}) and the
definition (\ref{K-def-via-J-X-T}) of the operator $\mathcal{K}_{i}$, we find
\begin{equation}
\exp\left( i\omega_{j}\mathcal{K}_{j}\right) \xi_{mKK_{3}}(g)\exp\left(
-i\omega_{j}\mathcal{K}_{j}\right)
=\sum_{K_{3}^{\prime}}\xi_{mKK_{3}^{\prime}}D_{K_{3}^{\prime}K_{3}}^{(K)}
\left( \exp\left( i\omega _{j}\mathcal{K}_{j}/2\right) \right) \,.
\end{equation}
This relation shows that the functional $\xi_{mKK_{3}}(g)$ belongs to the $K$
representation of the $SU(2)$ group generated by operators $\mathcal{K}_{i}$.

Our representation of the spin-flavor algebra in the space of functionals
$A_{B}(g)$ will allows us to work with the set of commuting operators 
$J^{2},J_{3},T^{2},T_{3},\mathcal{K}^{2}$ (\ref{JTK-set}). Obviously the
eigenstates of these operators are described by the functionals 
(\ref{Xi-Q-xi-1}).

\subsection{Factors $N_{c}^{\protect\nu_{B}}$}
\label{nu-B-subsection}

The definition (\ref{W-def-1}) of the functional $A_{B}(g)$ assumes the
separation of the power factor $N_{c}^{\nu _{B}}$. In the case of states 
$|B\rangle $ belonging to the same irreducible representation of the
large-$N_{c}$ spin-flavor symmetry group we have a common power $\nu _{B}$ for all
these states. However, in the general case parameter $\nu _{B}$ depends on
the state $|B\rangle $.

Creating an extra \emph{nonzero mode} excitation $\xi
_{m^{(j)}K^{(j)}K_{3}^{(j)}}(g)$ in Eq.~(\ref{Xi-Q-xi-1}), we shift $\nu
_{B} $ by $1/2$
\begin{equation}
\nu _{B}\rightarrow \nu _{B}+\frac{1}{2}\,.  \label{nu-B-shift}
\end{equation}
In order to derive this property, let us introduce the short notation 
$\mathcal{B}(g)$ for the baryon operator
\begin{equation}
\mathcal{B}(g)=N_{c}^{-N_{c}/2}\prod\limits_{c=1}^{N_{c}}\left[
\sum\limits_{fs}\int dyg_{fs}(y)\chi_{cfs}\left(\frac{y}{N_c}\right)\right]
\label{O-baryon-g-compact}
\end{equation}
made of fields $\chi_{cfs}(y)$ (\ref{chi-psi}).
Using Eqs. (\ref{BDA-def}), (\ref{Psi-B-prime-Psi-B}), and (\ref{Phi-def-1}),
we can rewrite Eq.~(\ref{W-def-1}) in the form
\begin{equation}
\langle 0|\mathcal{B}(g)|B\rangle =N_{c}^{\nu _{B}}A_{B}(g)\exp \left[
N_{c}W(g)\right] \,.
\label{O-g-B}
\end{equation}
Let $J$ be a color-singlet meson operator
\begin{equation}
J=\sum\limits_{c=1}^{N_{c}}\bar{\psi}_{c}\Gamma \psi _{c}\,,
\end{equation}
where $\Gamma $ is some spin-flavor matrix. Then by analogy with
Eq.~(\ref{O-g-B}) we can write
\begin{equation}
\langle 0|\mathcal{B}(g)J|B\rangle =N_{c}^{\nu _{BJ}}A_{BJ}(g)\exp \left[
N_{c}W(g)\right] \,.
\label{O-g-J-B}
\end{equation}
Note that the exponential behavior in Eqs. (\ref{O-g-B}) and (\ref{O-g-J-B})
is determined by the same universal functional $W(g)$, as was explained in
Ref. \cite{Pobylitsa-2004}. However, functional $A_{BJ}(g)$ and parameter
$\nu_{BJ}$ are different from $A_{B}(g)$ and $\nu_{B}$.

Obviously the presence of the operator $J$ in Eq.~(\ref{O-g-J-B}) leads
to an extra power of $N_{c}$. Therefore comparing Eqs. (\ref{O-g-B}) and
(\ref{O-g-J-B}), we find
\begin{equation}
\nu _{BJ}=\nu _{B}+1\,.  \label{nu-BJ-nu-B}
\end{equation}
The operator $J$ creates an extra meson $M$ in the state $|B\rangle $:
\begin{equation}
J|B\rangle \rightarrow |BM\rangle \,.
\end{equation}
Now we can write the large-$N_{c}$ asymptotic expression
(\ref{W-def-1}) for the state $|BM\rangle $:
\begin{equation}
\langle 0|\mathcal{B}(g)|BM\rangle =N_{c}^{\nu _{BM}}A_{BM}(g)\exp \left[
N_{c}W(g)\right] \,.  \label{O-g-BM}
\end{equation}
At large $N_{c}$ the connected part of the correlation function of two
currents $J$ has the behavior $\langle 0|JJ|0\rangle _{\mathrm{conn.}}\sim
N_{c}$. This allows us to estimate the matrix element $\langle 0|J|M\rangle
\sim \sqrt{N_{c}}$ corresponding to the intermediate meson state $M$.
Comparing this result with Eqs. (\ref{O-g-J-B}) and (\ref{O-g-BM}), we
conclude that $\nu _{BJ}=\nu _{BM}+1/2$. Together with Eq. (\ref{nu-BJ-nu-B})
this yields 
\begin{equation}
\nu _{BM}=\nu _{B}+\frac{1}{2}\,.  \label{nu-BM-shift}
\end{equation}%
This completes the derivation of the rule (\ref{nu-B-shift}).

We stress that the shift of $\nu_{B}$ (\ref{nu-B-shift}) is associated only
with the nonzero mode factors $\xi_{m^{(j)}K^{(j)}K_{3}^{(j)}}(g)$ in
Eq.~(\ref{Xi-Q-xi-1}). The zero mode factor $D_{T_{3}T_{3}^{\prime}}^{T}\left[
Q(g)\right] $ does not change $\nu_{B}$ since this factor is associated with
the transformations of the large-$N_{c}$ spin-flavor symmetry group.
These transformations connect different baryons belonging to the same
representation of the spin-flavor symmetry group. Parameter $\nu_B$
should be the same for all baryons within this multiplet.

\subsection{Lowest baryons and $N$-$\Delta$ relation}

The lowest $O(N_{c}^{-1})$ excited baryons are described by functionals $\Xi
_{T=J,T_{3}J_{3}}(g)$ (\ref{Xi-J-eq-T}). Combining Eqs. (\ref{Xi-AA-def}) and
(\ref{Xi-J-eq-T}), we obtain
\begin{equation}
A_{T_{3}J_{3}}^{T=J}(g)=\sqrt{2T+1}D_{T_{3}J_{3}}^{T}\left[ Q(g)\right]
A_{0}(g)\,.  \label{A-J-T}
\end{equation}
As was explained in Sec. \ref{Q-properties-section}, we use the short
notation $D^{T}$ implying the $SL(2,C)$ Wigner function $D^{(T,\dot 0)}$.
The $SL(2,C)$ Wigner function $D^{(J_1,{\dot J}_2)}(R)$
with $J_1=1/2$, ${\dot J}_2=0$ is
\begin{equation}
D_{mm^{\prime}}^{(1/2,\dot 0)}(R)=R_{mm^{\prime}}\,.
\end{equation}
Therefore, we obtain from Eq. (\ref{A-J-T}) for the nucleon case ($J=T=1/2$):
\begin{equation}
A_{T_{3}J_{3}}^{(N)}(g)=\sqrt{2}Q_{T_{3}J_{3}}(g)A_{0}(g)\,.
\label{A-N-Q-A-0}
\end{equation}
Taking into account Eq.~(\ref{det-Q-1}), we find
\begin{equation}
A_{0}(g)=\sqrt{\frac{1}{2}\det_{T_{3}J_{3}}A_{T_{3}J_{3}}^{(N)}(g)}\,,
\end{equation}
\begin{equation}
Q_{T_{3}J_{3}}(g)=\frac{A_{T_{3}J_{3}}^{(N)}(g)}{\sqrt{\det A^{(N)}(g)}}\,.
\end{equation}
Inserting these expressions into Eq.~(\ref{A-J-T}), we obtain
\begin{equation}
A_{T_{3}J_{3}}^{J=T}(g)=D_{T_{3}J_{3}}^{T}
\left( \frac{A^{(N)}(g)}{\sqrt{
\det A^{(N)}(g)}}\right) \sqrt{\left(T+\frac{1}{2}\right)
\det A^{(N)}(g)}\,.
\end{equation}

In particular, we find for the $\Delta $ resonance ($J=T=3/2$):
\begin{equation}
A_{T_{3}J_{3}}^{(\Delta )}(g)=D_{T_{3}J_{3}}^{3/2}
\left( \frac{A^{(N)}(g)}{
\sqrt{\det A^{(N)}(g)}}\right) \sqrt{2\det A^{(N)}(g)}\,.
\label{A-delta-A-N}
\end{equation}
Now let us consider the functionals $\Phi _{T_{3}J_{3}}^{(N)}(g)$ and 
$\Phi_{T_{3}J_{3}}^{(\Delta )}(g)$ (\ref{Phi-def-1})
describing the distribution amplitude of the nucleon and the $\Delta $.
The large-$N_{c}$ behavior of these
functionals is given by Eq.~(\ref{W-def-1}). We have $\nu _{\Delta }=\nu _{N}$
in Eq.~(\ref{W-def-1}) since the nucleon and the $\Delta $ resonance 
belong to the
same irreducible representation of the spin-flavor symmetry. Combining
Eqs. (\ref{W-def-1}) and (\ref{A-delta-A-N}), we find
\begin{equation}
\Phi _{T_{3}J_{3}}^{(\Delta )}(g)
=D_{T_{3}J_{3}}^{3/2}\left( \frac{\Phi
^{(N)}(g)}{\sqrt{\det \Phi ^{(N)}(g)}}\right) \sqrt{2\det \Phi ^{(N)}(g)}
\left[ 1+O(N_{c}^{-1})\right] \,.  \label{Phi-Delta-Phi-N}
\end{equation}
One should not try to use this relation for the derivation of a direct
expression for the distribution amplitude $\Psi ^{(\Delta)}(x_{1},x_{2},x_{3})
$ of the $\Delta $ resonance via the nucleon distribution amplitudes $\Psi
^{(N)}(x_{1},x_{2},x_{3})$ at finite $N_{c}=3$. Although at finite $N_{c}$ we
have the exact expression (\ref{Psi-B-via-Phi-B}) for the distribution
amplitude $\Psi ^{(\Delta)}$ ($\Psi ^{(N)}$) via the functional
$\Phi^{(\Delta)}(g)$ [$\Phi ^{(N)}(g)$], one cannot combine this exact relation with the
large-$N_{c}$ expression (\ref{Phi-Delta-Phi-N}), because the number
of variational derivatives in Eq.~(\ref{Psi-B-via-Phi-B}) grows as $N_{c}$.

Therefore practical applications of Eq.~(\ref{Phi-Delta-Phi-N}) should
deal directly with the functionals $\Phi^{(N)}(g)$, $\Phi^{(\Delta)}(g)$ which
can be considered as $g$ moments of the corresponding distribution
amplitudes.

\section{Soft-pion theorem}

\label{SP-theorem-section}

\subsection{Soft-pion theorem at finite $N_{c}$}

The soft-pion theorem for the distribution amplitude of the near-threshold
nucleon-pion state was derived and successfully applied to the description
of SLAC data on the near-threshold pion production in deeply virtual
scattering in Ref. \cite{PPS-01}. We want to derive the large-$N_{c}$
version of this theorem and to use it for the consistency check of the
representation (\ref{Xi-AA-def}), (\ref{Xi-Q-xi-1}) for the functional
$A_{B}(g)$.

We start from the general soft-pion theorem valid for any operator 
$\mathcal{O}$ at finite $N_{c}$, 
\begin{align}
& \langle 0|\mathcal{O}|N(T_{3}J_{3}),\pi _{a}(\mathbf{k})\rangle =-\frac{i}{
F_{\pi }}\langle 0|\left[ Q_{5a},\mathcal{O}\right] |N(T_{3}J_{3})\rangle  
\notag \\
& +\frac{ig_{A}}{2F_{\pi }}\sum\limits_{T_{3}^{\prime }J_{3}^{\prime
}n}\left( \tau _{a}\right) _{T_{3}^{\prime }T_{3}}\left( \tau _{n}\right)
_{J_{3}^{\prime }J_{3}}\frac{k^{n}}{E_{\pi }}\langle 0|\mathcal{O}
|N(T_{3}^{\prime }J_{3}^{\prime })\rangle \,.  \label{soft-pion-Nc-0}
\end{align}
Here $|N(T_{3}J_{3}),\pi _{a}(\mathbf{k})\rangle $ is a state containing

\begin{itemize}
\item nucleon $N(T_{3}J_{3})$ in the rest frame with spin $J_{3}$ and
isospin $T_{3}$,

\item soft pion $\pi _{a}(\mathbf{k})$ with small momentum $\mathbf{k}$,
energy $E_{\pi }=\sqrt{m_{\pi }^{2}+|\mathbf{k}|^{2}}$ and flavor index
$a=1,2,3$.
\end{itemize}

Note that for soft pions the difference between the \emph{in} and \emph{out}
scattering states can be ignored. The difference between the nucleon rest
frame and the c.m. frame is also not important in the soft-pion case.

The first term on the RHS of Eq.~(\ref{soft-pion-Nc-0}) contains the
commutator of the operator $\mathcal{O}$ with the axial charge 
\begin{equation}
Q_{5a}=\int d^{3}x\bar{q}\frac{\tau _{a}}{2}\gamma _{0}\gamma _{5}q\,,
\end{equation}
\begin{equation}
\left[ Q_{5a},q\right] =-\frac{1}{2}\gamma _{5}\tau _{a}q\,.
\label{Q5-q-commutator}
\end{equation}
The commutator term in Eq.~(\ref{soft-pion-Nc-0}) describes the $S$ wave of
the pion-nucleon state. This term does not vanish in the threshold limit
$\mathbf{k}\rightarrow 0$. It is accompanied by $1/F_{\pi }$, where 
\begin{equation}
F_{\pi }=93\,\mathrm{MeV}
\end{equation}
is the pion decay constant.

The second term on the RHS of Eq.~(\ref{soft-pion-Nc-0}) originates from the
pion-nucleon vertex and its contribution is proportional to the
Goldberger-Treiman ratio 
\begin{equation}
\frac{g_{A}}{F_{\pi }}=\frac{g_{\pi NN}}{M_{N}}\,.
\end{equation}
Here $g_{A}$ is the nucleon axial constant, $M_{N}$ is the nucleon mass and
$g_{\pi NN}$ is the pion-nucleon constant. The vertex also contains the spin
Pauli matrix $\left( \tau _{n}\right) _{J_{3}^{\prime }J_{3}}$ and the
isospin Pauli matrix $\left( \tau _{a}\right) _{T_{3}^{\prime }T_{3}}$.
Obviously the second term on the RHS of Eq.~(\ref{soft-pion-Nc-0}) describes
the $P$ wave. This term vanishes at the threshold $\mathbf{k}=\mathbf{0}$.

\subsection{Soft-pion theorem at large $N_{c}$}

The soft-pion theorem (\ref{soft-pion-Nc-0}) is valid only for those pions
whose energy $E_{\pi }$ is much smaller than the $\Delta $ production
threshold: 
\begin{equation}
E_{\pi }\ll M_{\Delta }-M_{N}\,.  \label{E-pi-small}
\end{equation}
In the limit $N_{c}\rightarrow \infty $ we have 
\begin{equation}
M_{\Delta }-M_{N}=O(N_{c}^{-1})\,.
\end{equation}
If we combine the condition (\ref{E-pi-small}) with the large-$N_{c}$ limit,
then the pion energy $E_{\pi }$ should approach zero much faster than
$1/N_{c}$.

In this section we want to study \emph{another} regime when the large-$N_{c}$
limit is taken \emph{before} the soft-pion limit: 
\begin{equation}
M_{\Delta }-M_{N}=O(N_{c}^{-1})\ll E_{\pi }\ll \Lambda_{\mathrm{QCD}}\,.
\label{N-c-E-pi-regime}
\end{equation}
In the real world, where $m_{\pi }<M_{\Delta }-M_{N}$, this regime is
certainly unphysical. However, our current aim is not the practical
applications of the soft-pion theorem. We want to use the large-$N_{c}$
version of the soft-pion theorem as a theoretical check of the consistency
of the factorized representation for the functional $A_{B}(g)$ 
(\ref{Xi-AA-def}), (\ref{Xi-Q-xi-single}).

In the region (\ref{N-c-E-pi-regime}) we must modify the soft-pion theorem
(\ref{soft-pion-Nc-0}) replacing the last term on the RHS by the sum over all 
$O(N_{c}^{-1})$ excited baryons which can appear in the intermediate state 
\begin{align}
& \langle 0|\mathcal{O}|B(T=J,T_{3}J_{3}),\pi _{a}(\mathbf{k})\rangle =-
\frac{i}{F_{\pi }}\langle 0|\left[ Q_{5a},\mathcal{O}\right]
|B(T=J,T_{3}J_{3})\rangle   \notag \\
& +\sum\limits_{T^{\prime }T_{3}^{\prime }J_{3}^{\prime }n}\frac{i}{2F_{\pi }
}G_{A}^{T^{\prime }T_{3}^{\prime }J_{3}^{\prime };TT_{3}J_{3};an}\frac{k^{n}
}{E_{\pi }}\langle 0|\mathcal{O}|B(T^{\prime }=J^{\prime },T_{3}^{\prime
}J_{3}^{\prime })\rangle \,.  \label{soft-pion-Nc-1}
\end{align}
Here $G_{A}^{T^{\prime }T_{3}^{\prime }J_{3}^{\prime };TT_{3}J_{3};an}$ is
the matrix element of the axial current corresponding to the transition
between the baryons $B(T=J,T_{3}J_{3})$ and $B(T^{\prime }=J^{\prime
},T_{3}^{\prime }J_{3}^{\prime })$. For example, in the nucleon case
($T=T^{\prime }=1/2$) we have 
\begin{equation}
G_{A}^{1/2,T_{3}^{\prime }J_{3}^{\prime };1/2,T_{3}J_{3};an}=g_{A}(\tau
_{a})_{T_{3}^{\prime }T_{3}}(\tau _{n})_{J_{3}^{\prime }J_{3}}
\label{G-A-nucleon}
\end{equation}
in agreement with the structure of the second term on the RHS of 
Eq.~(\ref{soft-pion-Nc-0}). The general large-$N_{c}$ expression for
$G_{A}^{T^{\prime }T_{3}^{\prime }J_{3}^{\prime };TT_{3}J_{3};an}$ can be
found in Appendix \ref{Pion-baryon-coupling-appendix}.

Let us take operator (\ref{O-baryon-g-compact}) for $\mathcal{O}$ in the
soft-pion theorem (\ref{soft-pion-Nc-1}): 
\begin{equation}
\mathcal{O}=\mathcal{B}(g)\,.
\end{equation}
According to Eq.~(\ref{Q5-q-commutator}) we have for this operator 
\begin{equation}
\left[ Q_{5a},\mathcal{B}(g)\right] =-\sum\limits_{f_{1}f_{2}s_{1}s_{2}}
\frac{1}{2}\int dyg_{f_{1}s_{1}}(y)\left( \tau _{a}\right)
_{f_{1}f_{2}}\left( \gamma _{5}\right) _{s_{1}s_{2}}\frac{\delta }{\delta
g_{f_{2}s_{2}}(y)}\mathcal{B}(g)\,.
\end{equation}
Now we find from Eq.~(\ref{soft-pion-Nc-1}) 
\begin{align}
& \langle 0|\mathcal{B}(g)|B(T=J,T_{3}J_{3}),\pi _{a}(\mathbf{k})\rangle  
\notag \\
& =\frac{i}{2F_{\pi }}\sum\limits_{f_{1}f_{2}s_{1}s_{2}}\int
dyg_{f_{1}s_{1}}(y)\left( \tau _{a}\right) _{f_{1}f_{2}}\left( \gamma
_{5}\right) _{s_{1}s_{2}}\frac{\delta }{\delta g_{f_{2}s_{2}}(y)}\langle 0|
\mathcal{B}(g)|B(T=J,T_{3}J_{3})\rangle   \notag \\
& +\sum\limits_{T^{\prime }T_{3}^{\prime }J_{3}^{\prime }n}\frac{i}{2F_{\pi }
}G_{A}^{T^{\prime }T_{3}^{\prime }J_{3}^{\prime };TT_{3}J_{3};an}\frac{k^{n}
}{E_{\pi }}\langle 0|\mathcal{B}(g)|B(T^{\prime }=J^{\prime },T_{3}^{\prime
}J_{3}^{\prime })\rangle \,.  \label{O-g-soft-theorem}
\end{align}

\subsection{Functional $A_{B\protect\pi }(g)$ for pion-baryon states}

Using Eq.~(\ref{O-g-soft-theorem}), we can compute the functional $A_{B\pi
}(g)$ corresponding to the large-$N_{c}$ distribution amplitude of the
pion-baryon scattering state with a soft pion. According to 
Eqs. (\ref{W-def-1}) and (\ref{nu-BM-shift}) we expect that at large $N_{c}$ 
\begin{equation}
\langle 0|\mathcal{B}(g)|B(T=J,T_{3}J_{3})\rangle =N_{c}^{\nu
_{B}}A_{B(T=J,T_{3}J_{3})}(g)\exp \left[ N_{c}W(g)\right] \,,
\end{equation}
\begin{equation}
\langle 0|\mathcal{B}(g)|B(T=J,T_{3}J_{3}),\pi _{a}(\mathbf{k})\rangle
=N_{c}^{\nu _{B}+\frac{1}{2}}A_{B(T=J,T_{3}J_{3}),\pi _{a}(\mathbf{k}
)}(g)\exp \left[ N_{c}W(g)\right] \,.
\end{equation}
Inserting this ansatz into Eq.~(\ref{O-g-soft-theorem}), we obtain in the
leading order of the $1/N_{c}$ expansion 
\begin{align}
N_{c}^{1/2}& A_{B(T=J,T_{3}J_{3}),\pi _{a}(\mathbf{k})}(g)=\frac{iN_{c}}{
2F_{\pi }}\left( g\tau _{a}\gamma _{5}\cdot \frac{\delta W}{\delta g}\right)
A_{B(T=J,T_{3}J_{3})}(g)  \notag \\
& +\sum\limits_{T^{\prime }T_{3}^{\prime }J_{3}^{\prime }n}\frac{i}{2F_{\pi }
}G_{A}^{T^{\prime }T_{3}^{\prime }J_{3}^{\prime };TT_{3}J_{3};an}\frac{k^{n}
}{E_{\pi }}A_{B(T^{\prime }=J^{\prime },T_{3}^{\prime }J_{3}^{\prime
})}(g)\,\,.  \label{soft-pion-Nc-2}
\end{align}
Here we use the short notation 
\begin{equation}
\left( g\tau _{a}\gamma _{5}\cdot \frac{\delta W}{\delta g}\right)
=\sum\limits_{f_{1}f_{2}s_{1}s_{2}}\int dyg_{f_{1}s_{1}}(y)\left( \tau
_{a}\right) _{f_{1}f_{2}}\left( \gamma _{5}\right) _{s_{1}s_{2}}\frac{\delta
W(g)}{\delta g_{f_{2}s_{2}}(y)}\,.
\end{equation}

Substituting Eq.~(\ref{A-J-T}) into Eq.~(\ref{soft-pion-Nc-2}), we find 
\begin{align}
& N_{c}^{1/2}A_{B(T=J,T_{3}J_{3}),\pi _{a}(\mathbf{k})}(g)  \notag \\
& =\frac{i}{2F_{\pi }}A_{0}(g)\left[ N_{c}\sqrt{2T+1}\left( g\tau _{a}\gamma
_{5}\cdot \frac{\delta W}{\delta g}\right) D_{T_{3}J_{3}}^{T}\left(
Q(g)\right) \right.   \notag \\
& +\sum\limits_{T^{\prime }T_{3}^{\prime }J_{3}^{\prime }n}\left. \frac{k^{n}
}{E_{\pi }}G_{A}^{T^{\prime }T_{3}^{\prime }J_{3}^{\prime };TT_{3}J_{3};an}
\sqrt{2T^{\prime }+1}D_{T_{3}^{\prime }J_{3}^{\prime }}^{T^{\prime }}\left(
Q(g)\right) \right] \,.  \label{soft-pion-Nc-3}
\end{align}
Now we use identity (\ref{G-D}) from Appendix \ref{Pion-baryon-coupling-appendix}: 
\begin{equation}
\sum\limits_{T^{\prime }T_{3}^{\prime }J_{3}^{\prime }}G_{A}^{T^{\prime
}T_{3}^{\prime }J_{3}^{\prime };TT_{3}J_{3};an}\sqrt{2T^{\prime }+1}
D_{T_{3}^{\prime }J_{3}^{\prime }}^{T^{\prime }}\left( Q(g)\right) =-3g_{A}
\sqrt{2T+1}D_{an}^{\mathrm{vec}}\left( i\tau _{2}Q(g)\right)
D_{T_{3}J_{3}}^{T}\left( Q(g)\right) \,.  \label{G-DD}
\end{equation}
Function $D_{an}^{\mathrm{vec}}$ is defined in Eq. (\ref{D-vec-def}).
Inserting Eq. (\ref{G-DD}) into Eq.~(\ref{soft-pion-Nc-3}), we arrive at 
\begin{align}
A_{B(T^{B}=J^{B},T_{3}^{B}J_{3}^{B}),\pi _{a}(\mathbf{k})}(g)& =\frac{
iN_{c}^{1/2}}{2F_{\pi }}A_{0}(g)D_{T_{3}^{B}J_{3}^{B}}^{T^{B}}\left(
Q(g)\right) \sqrt{2T^{B}+1}  \notag \\
& \times \left[ \left( g\tau _{a}\gamma _{5}\cdot \frac{\delta W}{\delta g}
\right) -3\frac{g_{A}}{N_{c}}\frac{k^{n}}{E_{\pi }}D_{an}^{\mathrm{vec}
}\left( i\tau _{2}Q(g)\right) \right] \,.  \label{soft-pion-Nc-4}
\end{align}
In this equation we have changed the notation: 
\begin{equation}
T\rightarrow T^{B},J\rightarrow J^{B}
\end{equation}
in order to emphasize that we deal with the baryon quantum numbers and not
with the total isospin and angular momentum of the baryon-pion state.

Note that at large $N_{c}$ 
\begin{equation}
F_{\pi }=O(N_{c}^{1/2})\,,\quad g_{A}=O(N_{c})\,.
\end{equation}
Therefore the ratios $N_{c}^{1/2}/F_{\pi }$ and $g_{A}/N_{c}$ appearing in
Eq.~(\ref{soft-pion-Nc-4}) have finite large-$N_{c}$ limits. We can rewrite
Eq. (\ref{soft-pion-Nc-4}) in terms of the functional $\Xi (g)$
(\ref{Xi-AA-def}): 
\begin{align}
\Xi _{B(T^{B}=J^{B},T_{3}^{B}J_{3}^{B}),\pi _{a}(\mathbf{k})}(g)& =\frac{
iN_{c}^{1/2}}{2F_{\pi }}D_{T_{3}^{B}J_{3}^{B}}^{T^{B}}\left( Q(g)\right) 
\sqrt{2T^{B}+1}  \notag \\
& \times \left[ \left( g\tau _{a}\gamma _{5}\cdot \frac{\delta W}{\delta g}
\right) -3\frac{g_{A}}{N_{c}}\frac{k^{n}}{E_{\pi }}D_{an}^{\mathrm{vec}
}\left( i\tau _{2}Q(g)\right) \right] \,.  \label{Xi-B-soft-pi-res}
\end{align}

\subsection{Consistency of the large-$N_{c}$ factorization with the
soft-pion theorem}

Our aim is to show that the general structure (\ref{Xi-Q-xi-1}) of the
functional $\Xi (g)$ is consistent with the soft-pion theorem. The main part
of this work has been done. We have derived the factorized expression
(\ref{Xi-B-soft-pi-res}) for the functional $\Xi _{B\pi }(g)$ describing the
baryon-pion state with a soft pion. Our result (\ref{Xi-B-soft-pi-res}) is
written for the $B\pi $ states characterized by separate quantum numbers
$T^{B}=J^{B},T_{3}^{B}J_{3}^{B}$ for the baryon and $a,\mathbf{k}$ for the
pion: 
\begin{equation}
|B\left( T^{B}=J^{B},T_{3}^{B}J_{3}^{B}\right) ;\pi _{a}(\mathbf{k})\rangle
\,.  \label{B-pi-basis}
\end{equation}
However, the general expression (\ref{Xi-Q-xi-1}) for the functional $\Xi (g)
$ is formulated for the eigenstates 
\begin{equation}
|TT_{3}JJ_{3}KL,k\rangle \,,\quad k\equiv |\mathbf{k}|  \label{TJK-basis}
\end{equation}
of the set of commuting operators (\ref{JTK-set}). In addition to quantum
numbers (\ref{JTK-set}), here we also have the orbital angular momentum $L$.
Since the pion has no spin, the total angular momentum is 
\begin{equation}
\mathbf{J}=\mathbf{J}^{B}+\mathbf{L}\,.
\end{equation}
Strictly speaking, neither $J^{B}$ nor $L$ are conserved in the pion-baryon
scattering. The precise definition of states (\ref{TJK-basis}) should be
formulated in terms of \emph{in}- and \emph{out}-states, $S$-matrix, etc.,
as is usually done in the large-$N_{c}$ approach to the meson-baryon
scattering \cite{HEHW-84,MK-85,MP-85,Mattis-89,MM-88,MB-89,DPP-88}. However,
for the soft pions the difference between the \emph{in}- and
\emph{out}-states can be neglected.

The transition from the basis of states (\ref{B-pi-basis}) to the basis
(\ref{TJK-basis}) is a matter of standard manipulations with the angular momentum
which are described in Appendix \ref{Soft-pion-factors-appendix}. As a
result of this change of the basis, one passes from the functionals $\Xi
_{B(T^{B}=J^{B},T_{3}^{B}J_{3}^{B}),\pi _{a}(\mathbf{k})}(g)$
(\ref{Xi-B-soft-pi-res}) to the functionals $\Xi _{TT_{3}JJ_{3}B,\pi (k,L)}^{K}(g)
$. One can see directly from Eq.~(\ref{Xi-B-soft-pi-res}) that this
amplitude contains only the waves with $L=0$ and $L=1$. It is also easy to
check (see Appendix \ref{Soft-pion-factors-appendix}) that the $L=0$
component corresponds to the value $K=1$ whereas the $L=1$ contribution
corresponds to $K=0$. As a result, in the soft-pion approximation
(\ref{Xi-B-soft-pi-res}) we have only two nonzero components of $\Xi
_{TT_{3}JJ_{3}B,\pi (k,L)}^{K}(g)$: 
\begin{equation}
\Xi _{TT_{3}JJ_{3}B,\pi (k,L=1)}^{K=0}(g)=-i\sqrt{\pi }\frac{N_{c}^{1/2}}{
F_{\pi }}\left[ \delta _{TJ}\sqrt{2T+1}D_{T_{3}J_{3}}^{T}\left( Q(g)\right) 
\frac{3g_{A}}{N_{c}}\frac{k}{E_{\pi }}\right] \,,  \label{Xi-soft-K-0}
\end{equation}
\begin{gather}
\Xi _{TT_{3}JJ_{3}B,\pi (k,L=0)}^{K=1}(g)=i\sqrt{\pi }\frac{N_{c}^{1/2}}{
F_{\pi }}  \notag \\
\times \left[ \sqrt{(2T+1)(2J+1)}\delta _{JJ^{B}}\sum\limits_{T_{3}^{\prime
}K_{3}}(-1)^{T+1+J_{3}}\left( 
\begin{array}{ccc}
T & 1 & J \\ 
T_{3}^{\prime } & K_{3} & -J_{3}
\end{array}
\right) D_{T_{3}T_{3}^{\prime }}^{T}\left( Q(g)\right) \zeta _{K_{3}}(g)
\right] \,.  \label{Xi-soft-K-1}
\end{gather}
The last equation contains functional $\zeta _{K_{3}}(g)$ whose expression
via the universal functional $W(g)$ is given by Eq.~(\ref{chi-via-Q-W}).

The structure of these results completely agrees with the general
representation (\ref{Xi-Q-xi-single}). Note that the factors 
$\xi_{mKK_{3}}(g)$ of Eq.~(\ref{Xi-Q-xi-single}) are now associated with the
soft pions. In this case the role of index $m$ in $\xi_{mKK_{3}}(g)$ is
played by two pion parameters 
\begin{equation}
m=\{L,k\}\,.
\end{equation}
Comparing Eqs. (\ref{Xi-soft-K-0}) and (\ref{Xi-soft-K-1}) with
Eq.~(\ref{Xi-Q-xi-single}), we find 
\begin{equation}
\xi_{\{L=0,k\},K=1,K_{3}}(g)=i\sqrt{\pi}\frac{N_{c}^{1/2}}{F_{\pi}}\zeta
_{K_{3}}(g)\,,  \label{xi-L0-K1}
\end{equation}
\begin{equation}
\xi_{\{L=1,k\},K=0,K_{3}=0}(g)=-i\sqrt{\pi}\frac{N_{c}^{1/2}}{F_{\pi}}\frac{
3g_{A}}{N_{c}}\frac{k}{E_{\pi}}\,.  \label{xi-L1-K0}
\end{equation}

Note that the functional $\xi _{\{L=1,k\},K=0,K_{3}}(g)$ is $g$ independent.
In principle, this $g$ independence is an artefact of the soft-pion limit.
Nevertheless this independence is a good illustration of the warning made in
Sec.~\ref{Functional-AB-representation-subsection}: The interpretation of
functionals $A_{B}(g)$ as a special representation for baryon-meson states 
$|B\rangle $ is not quite correct since different states $|B\rangle $ can be
mapped to identical functionals $A_{B}(g)$.

\subsection{Lessons from the soft-pion theorem}

The result of our analysis of the soft-pion theorem is represented by the
expressions (\ref{xi-L0-K1}), (\ref{xi-L1-K0}) for the two nonvanishing
functionals $\xi_{\{L,k\}KK_{3}}(g)$ describing the $L=0$ and $L=1$ waves of
the soft-pion scattering states. The result of the calculation is not of the
primary importance for us. What is really significant is that the physics of
soft pions is compatible with the general representation (\ref{Xi-AA-def}),
(\ref{Xi-Q-xi-single}) for the functionals $A_{B}(g)$.

In our analysis we have concentrated on the case of baryon-pion states with
one pion. But the factorized structure of the expression
(\ref{Xi-B-soft-pi-res}) allows for a straightforward generalization to the
multipion case so that the representation (\ref{Xi-AA-def}),
(\ref{Xi-Q-xi-single}) for $A_{B}(g)$ can be checked also for multipion states.

In this paper, the consistency of the representation (\ref{Xi-Q-xi-single})
was checked using two methods:

\begin{itemize}
\item spin-flavor symmetry,

\item soft-pion theorem.
\end{itemize}

It is well known that there is a deep connection between the two approaches.
The spin-flavor symmetry has an elegant derivation
\cite{Bardakci-84,GS-84,DM-93} based on the analysis of the soft pion-baryon
scattering (strictly speaking, the chiral limit and the existence of the
Goldstone bosons are not necessary for the spin-flavor symmetry, this
symmetry is a consequence of the large-$N_{c}$ limit only). Our analysis of
the soft-pion theorem for the functional $A_{B}(g)$ and the traditional
derivation \cite{Bardakci-84,GS-84,DM-93} of the spin-flavor symmetry from
the consistency condition for the pion-baryon scattering have many common
points.

\section{Conclusions}

The large-$N_{c}$ description of the distribution amplitude of baryon $B$ in
terms of the generating functional $\Phi_{B}(g)$ is based on the important
property of universality. The exponential behavior $\Phi_{B}(g)\sim\exp\left[
N_{c}W(g)\right] $ is governed by the same functional $W(g)$ for all baryons
(and baryon-meson scattering states) with the excitation energies 
$O(N_{c}^{-1})$ and $O(N_{c}^{0})$.

However, if one goes beyond the exponential accuracy and considers the
preexponential factors $A_{B}(g)$, one loses the universality: the
functional $A_{B}(g)$ depends on the state $B$. Although we do not know the
explicit expression for $A_{B}(g)$, we have found that $A_{B}(g)$ has a simple
factorized structure which can be described in terms of elementary excitations
and zero mode factors.

Some evidence for the factorization of $A_{B}(g)$ was observed
earlier \cite{Pobylitsa-2004} in the context of the evolution equation and
the asymptotic limit of large scales $\mu\rightarrow\infty$. This work
provides much more solid arguments based on the spin-flavor symmetry. We
have also demonstrated the consistency of the large-$N_{c}$ factorization of 
$A_{B}(g)$ with the soft-pion theorem.

Combining the universality of the exponential large-$N_{c}$ behavior and the
factorized structure of $A_{B}(g)$, we have derived a simple relation
between the generating functionals describing the distribution amplitudes of
the nucleon and the $\Delta$ resonance.

\acknowledgments I am grateful to Ya.~I.~Azimov, V.~Yu.~ Petrov and
M.~V.~Polyakov for interesting discussions.
This work was supported by DFG and BMBF.

\appendix

\section{Wigner functions}
\label{Wigner-functions-appendix}

The Wigner functions $D_{m_{1}m_{2}}^{j}(R)$ for the $SU(2)$ group have the well
known properties:
\begin{equation}
D_{m_{1}m_{2}}^{j}(R_{1}R_{2})=\sum
\limits_{m=-j}^{j}D_{m_{1}m}^{j}(R_{1})D_{mm_{2}}^{j}(R_{2}),
\label{D-property-1}
\end{equation}
\begin{equation}
\left[ D_{mm^{\prime}}^{j}(R)\right] ^{\ast}=(-1)^{m-m^{\prime}}D_{-m,-m^{
\prime}}^{j}(R)\,,
\end{equation}
\begin{equation}
D_{mm^{\prime}}^{j}(i\tau_{2})=(-1)^{j-m}\delta_{m,-m^{\prime}}\,,
\label{D-i-tau-2}
\end{equation}
\begin{equation}
D_{m_{1}m_{2}}^{j}(R^{\mathrm{tr}})=D_{m_{2}m_{1}}^{j}(R)\,.
\label{D-R-transposed}
\end{equation}
Here $R^{\mathrm{tr}}$ stands for the transposed matrix.

The $SU(2)$ invariant measure $dR$ is normalized by the condition
\begin{equation}
\int dR=1\,.
\end{equation}
Then
\begin{equation}
\int dR\left[ D_{m_{1}m_{2}}^{j}(R)\right] ^{\ast}D_{m_{1}^{\prime}m_{2}^{
\prime}}^{j^{\prime}}(R)=\frac{1}{2j+1}\delta_{jj^{\prime}}
\delta_{m_{1}m_{1}^{\prime}}\delta_{m_{2}m_{2}^{\prime}}\,.
\label{D-property-2}
\end{equation}
The integral of the product of three $D$ functions can be expressed via
$3j$ symbols:
\begin{equation}
\int
dRD_{m_{1}^{\prime}m_{1}}^{j_{1}}(R)D_{m_{2}^{
\prime}m_{2}}^{j_{2}}(R)D_{m_{3}^{\prime}m_{3}}^{j_{3}}(R)=\left( 
\begin{array}{ccc}
j_{1} & j_{2} & j_{3} \\ 
m_{1}^{\prime} & m_{2}^{\prime} & m_{3}^{\prime}
\end{array}
\right) \left( 
\begin{array}{ccc}
j_{1} & j_{2} & j_{3} \\ 
m_{1} & m_{2} & m_{3}
\end{array}
\right) \,.  \label{int-3D-3j}
\end{equation}

The standard Wigner function $D_{mm^{\prime }}^{1}(R)$ corresponding to the
spin $j=1$ is equivalent to the vector representation $D_{ab}^{\mathrm{vec}
}(R)$:
\begin{equation}
D_{ab}^{\mathrm{vec}}(R)=\frac{1}{2}\mathrm{Tr}\left( \tau _{a}R\tau
_{b}R^{-1}\right) \,,  \label{D-vec-def}
\end{equation}
\begin{equation}
D_{mm^{\prime }}^{1}(R)=\sum\limits_{ab}O_{am}^{\ast }O_{bm^{\prime}}
D_{ab}^{\mathrm{vec}}(R)\,.  \label{D1-via-D-vec}
\end{equation}
The matrix $O_{am}$ is given in Table~\ref{O-at-def}. This matrix
performs the unitary transformation from the spin-one states labeled by the
momentum projection $m=0,\pm 1$ to the basis with vector indices $a=1,2,3$.
Matrix $O_{am}$ has the properties
\begin{equation}
OO^{+}=1,
\end{equation}
\begin{equation}
O_{am}^{\ast }=(-1)^{m+1}O_{a,-m}\,\,.  \label{O-conjugate}
\end{equation}

\begin{table}
\begin{tabular}{|cc|ccc|}
\hline
& $m$ & $-1$ & 0 & 1 \\ 
$a$ &  &  &  &  \\ \hline
$1$ &  & $i/\sqrt{2}$ & 0 & $-i/\sqrt{2}$ \\ 
$2$ &  & $1/\sqrt{2}$ & 0 & $1/\sqrt{2}$ \\ 
$3$ &  & $0$ & $i$ & $0$ \\ \hline
\end{tabular}
\caption{Matrix $O_{am}$.}
\label{O-at-def}
\end{table}

The spherical functions $Y_{lm}(\mathbf{n})$ with $l=0$ and $l=1$ have the
form
\begin{equation}
Y_{00}(\mathbf{n})=\frac{1}{\sqrt{4\pi }}\,,  \label{Y00-explicit}
\end{equation}
\begin{equation}
Y_{1m}(\mathbf{n})=\sqrt{\frac{3}{4\pi }}n^{i}O_{im}\,.
\label{Y1-m-explicit}
\end{equation}
The connection between the Clebsh-Gordan coefficients and $3j$ symbols is
\begin{equation}
C_{j_{1}m_{1}j_{2}m_{2}}^{jm}=(-1)^{j_{1}-j_{2}+m}\sqrt{2j+1}\left( 
\begin{array}{ccc}
j_{1} & j_{2} & j \\ 
m_{1} & m_{2} & -m
\end{array}
\right) \,.  \label{C-3j}
\end{equation}

\section{Axial baryon coupling constants at large $N_{c}$}
\label{Pion-baryon-coupling-appendix}

At large $N_{c}$ the axial baryon couplings $G_{A}^{T^{\prime }T_{3}^{\prime
}J_{3}^{\prime };TT_{3}J_{3};an}$ appearing in Eq.~(\ref{soft-pion-Nc-1})
can be expressed in terms of $3j$ symbols:
\begin{align}
G_{A}^{T^{\prime }T_{3}^{\prime }J_{3}^{\prime };TT_{3}J_{3};an}&
=-3g_{A}(-1)^{J+J_{3}^{\prime }}(-1)^{T+T_{3}^{\prime }}\sqrt{\left(
2T^{\prime }+1\right) \left( 2T+1\right) }  \notag \\
& \times \sum\limits_{tm}O_{nm}^{\ast }O_{at}^{\ast }\left( 
\begin{array}{ccc}
T & 1 & T^{\prime } \\ 
T_{3} & t & -T_{3}^{\prime }
\end{array}
\right) \left( 
\begin{array}{ccc}
J & 1 & J^{\prime } \\ 
J_{3} & m & -J_{3}^{\prime }
\end{array}
\right) \,.  \label{G-A-3j}
\end{align}
Here
\begin{equation}
J^{\prime }=T^{\prime }\,,\quad J=T\,,
\end{equation}
$g_{A}$ is the axial constant of nucleon, and matrix $O_{nm}$ is defined in
Table~\ref{O-at-def}.

In the context of the spin-flavor symmetry of large-$N_{c}$ QCD, 
Eq.~(\ref{G-A-3j}) is discussed in Refs. \cite{DM-93,DJM-94}. In the
Skyrme model and
other chiral models expression (\ref{G-A-3j}) appears via the integral over
the $SU(2)$ matrices $R$ \cite{Adkins-83,Manohar-84}
\begin{equation}
G_{A}^{T^{\prime }T_{3}^{\prime }J_{3}^{\prime };TT_{3}J_{3};an}=-3g_{A}\int
dR\left[ \psi _{T_{3}^{\prime }J_{3}^{\prime }}^{T^{\prime }}(R)\right]
^{\ast }D_{an}^{\mathrm{vec}}(R)\psi _{T_{3}J_{3}}^{T}(R)\,\,,
\label{G-A-g-A}
\end{equation}
where $\psi_{T_{3}J_{3}}^{T}(R)$ are the rotational wave functions of the
soliton
\begin{equation}
\psi _{T_{3}J_{3}}^{T=J}(R)=(-1)^{T+T_{3}}\sqrt{2T+1}
D_{-T_{3},J_{3}}^{T=J}(R)\,,
\end{equation}
and function $D_{an}^{\mathrm{vec}}(R)$ is defined by Eq.~(\ref{D-vec-def}).
The normalization coefficient in Eq. (\ref{G-A-3j}) is
chosen so that in the nucleon case $T=T^{\prime }=1/2$ one obtains
Eq.~(\ref{G-A-nucleon}).

It is straightforward to show that
\begin{equation}
\sum\limits_{T^{\prime }J_{3}^{\prime }T_{3}^{\prime }}G_{A}^{T^{\prime
}T_{3}^{\prime }J_{3}^{\prime };TT_{3}J_{3};an}\sqrt{2T^{\prime }+1}
D_{T_{3}^{\prime }J_{3}^{\prime }}^{T^{\prime }}(R)=-3g_{A}\sqrt{2T+1}
D_{an}^{\mathrm{vec}}(i\tau _{2}R)D_{T_{3}J_{3}}^{T}(R)\,.  \label{G-D}
\end{equation}
This equality also holds for arbitrary $SL(2,C)$ matrices $R$ if one
replaces the $SU(2)$ Wigner functions $D^{T}$ by their $SL(2,C)$
generalization $D^{(T,\dot 0)}$.

\section{Soft-pion factors}

\label{Soft-pion-factors-appendix}

In this appendix, we construct the baryon-meson states (\ref{TJK-basis})
from the states (\ref{B-pi-basis}) and compute the corresponding
functionals $\Xi_{TT_{3}JJ_{3}B,\pi(k,L)}^{K}(g)$
(\ref{Xi-soft-K-0}) and (\ref{Xi-soft-K-1}).
As a first step we build the
eigenstates of operators
\begin{equation}
T^{2},T_{3},\,J^{2},J_{3},\,L^{2},\left( J^{B}\right) ^{2}=\left(
T^{B}\right) ^{2}\,,
\end{equation}
where $T$ is the total isospin, $J$ is the total angular momentum, and $L$
is the orbital momentum. In order to avoid confusion, we mark the baryon
quantum numbers  $J^{B}=T^{B}$ with the label $B$. The standard
rules of the addition of angular momenta yield:
\begin{align}
|TT_{3}JJ_{3},B(T^{B}& =J^{B}),\pi (k,L)\rangle  \notag \\
&
=\sum\limits_{T_{3}^{B}t}\sum
\limits_{J_{3}^{B}m}C_{T^{B}T_{3}^{B},1t}^{TT_{3}}C_{J^{B}J_{3}^{B},Lm}^{JJ_{3}}\sum\limits_{a}
\int d^{2}{\bf n}Y_{Lm}
(\mathbf{n})O_{at}|B(T^{B}=J^{B},T_{3}^{B}J_{3}^{B}),\pi_a (k\mathbf{n})\rangle \,.  
\label{Bpi-TTJJ}
\end{align}
Here $O_{at}$ is the matrix from Table~\ref{O-at-def}. This matrix  converts the states $|\pi
_{a}\rangle $ with indices $a=1,2,3$ (associated with the Pauli matrices 
$\tau_{a}$) to the eigenstates of the isospin $t=0,\pm 1$.
The RHS of Eq.~(\ref{Bpi-TTJJ}) contains the integration over the direction
$\bf n$ of the pion momentum $k{\bf n}$.
Combining  Eq.~(\ref{Bpi-TTJJ}) with Eq.~(\ref{soft-pion-Nc-4}), we find
\begin{align}
& A_{TT_{3}JJ_{3}B(T^{B}=J^{B}),\pi (k,L)}(g)=\int d^{2}{\bf n}
\sum\limits_{T_{3}^{B}t}\sum\limits_{J_{3}^{B}m}
Y_{Lm}(\mathbf{n})
C_{T^{B}T_{3}^{B},1t}^{TT_{3}}C_{J^{B}J_{3}^{B},Lm}^{JJ_{3}}
\frac{iN_{c}^{1/2}}{2F_{\pi }}A_{0}(g)D_{T_{3}^{B}J_{3}^{B}}^{T^{B}}
\left(Q(g)\right)  \notag \\
& \times \sqrt{2T^{B}+1}\left[ \eta _{t}(g)-3\frac{g_{A}}{N_{c}}\frac{k}{
E_{\pi }}\sum\limits_{aj}n^{j}O_{at}D_{aj}^{\mathrm{vec}}
\left( i\tau_{2}Q(g)\right) \right] \,,
\end{align}
where
\begin{equation}
\eta _{t}(g)=\sum\limits_{a}O_{at}\left[ g\tau_{a}\gamma_{5}\cdot \frac{
\delta W(g)}{\delta g}\right] \,.  \label{eta-t-def}
\end{equation}
Next we integrate over $\mathbf{n}$ using expressions
(\ref{Y00-explicit}) and (\ref{Y1-m-explicit}) for $Y_{Lm}(\mathbf{n})$.
Applying Eqs. (\ref{D-i-tau-2}), (\ref{D1-via-D-vec}), and (\ref{O-conjugate}),
we find
\begin{gather}
A_{TT_{3}JJ_{3}B(T^{B}=J^{B}),\pi (k,L)}(g)=i\sqrt{\pi }\frac{N_{c}^{1/2}}{
F_{\pi }}A_{0}(g)  \notag \\
\times \left[ \delta _{L0}\sqrt{2J+1}\delta
_{JJ^{B}}\sum
\limits_{T_{3}^{B}t}C_{T^{B}T_{3}^{B},1t}^{TT_{3}}D_{T_{3}^{B}J_{3}}^{T^{B}}
\left( Q(g)\right) \eta _{t}(g)-\sqrt{2T^{B}+1}\delta _{L1}\delta
_{TJ}D_{T_{3}J_{3}}^{J}\left( Q(g)\right) \sqrt{3}\frac{g_{A}}{N_{c}}\frac{k
}{E_{\pi }}\right] \,.  \label{A-calc-1}
\end{gather}
Let us introduce the notation
\begin{equation}
\zeta _{K_{3}}(g)=(-1)^{1+K_{3}}\sum\limits_{t}D_{-K_{3},t}^{1}
\left(Q^{-1}(g)\right) \eta _{t}(g)\,.  \label{chi-via-eta-def}
\end{equation}
Inserting Eq.~(\ref{eta-t-def}) into Eq.~(\ref{chi-via-eta-def}), we obtain
\begin{equation}
\zeta _{K_{3}}(g)=(-1)^{1+K_{3}}\sum\limits_{t}D_{-K_{3},t}^{1}
\left(Q^{-1}(g)\right) \sum\limits_{a}O_{at}\left[ g\tau _{a}\gamma_{5}\cdot 
\frac{\delta W(g)}{\delta g}\right] \,.  \label{chi-via-Q-W}
\end{equation}
Now we find from Eq.~(\ref{A-calc-1}) using Eq.~(\ref{chi-via-eta-def})
\begin{gather}
A_{TT_{3}JJ_{3}B(T^{B}=J^{B}),\pi (k,L)}(g)=i\sqrt{\pi }\frac{N_{c}^{1/2}}{
F_{\pi }}A_{0}(g)  \notag \\
\times \left[ \delta _{L0}\sqrt{2J+1}\delta
_{JJ^{B}}\sum\limits_{T_{3}^{\prime
}K_{3}}(-1)^{1-K_{3}}C_{T^{B}J_{3},1,-K_{3}}^{TT_{3}^{\prime
}}D_{T_{3}T_{3}^{\prime }}^{T}\left( Q(g)\right) \zeta _{K_{3}}(g)
-\delta_{L1}\delta _{TJ}\sqrt{2T^{B}+1}D_{T_{3}J_{3}}^{J}\left( Q(g)\right) \sqrt{3}
\frac{g_{A}}{N_{c}}\frac{k}{E_{\pi }}\right] \,.
\label{A-K-decomposition}
\end{gather}
In terms of the functional (\ref{Xi-AA-def})
\begin{equation}
\Xi _{TT_{3}JJ_{3}B(T^{B}=J^{B}),\pi (k,L)}(g)=\left[ A_{0}(g)\right]
^{-1}A_{TT_{3}JJ_{3}B(T^{B}=J^{B}),\pi (k,L)}(g)
\end{equation}
we obtain from Eq.~(\ref{A-K-decomposition})
\begin{gather}
\Xi _{TT_{3}JJ_{3}B(T^{B}=J^{B}),\pi (k,L)}(g)=i\sqrt{\pi }\frac{N_{c}^{1/2}
}{F_{\pi }}  \notag \\
\times \Biggl[ \delta _{L0}\sqrt{(2T+1)(2J+1)}\delta
_{JJ^{B}}\sum\limits_{T_{3}^{\prime }K_{3}}(-1)^{J+J_{3}}\left( 
\begin{array}{ccc}
T & 1 & J \\ 
T_{3}^{\prime } & K_{3} & -J_{3}
\end{array}
\right) D_{T_{3}T_{3}^{\prime }}^{T}\left( Q(g)\right) \zeta
_{K_{3}}(g)
\notag \\
-\delta _{L1}\delta _{TJ}\sqrt{2T^{B}+1}D_{T_{3}J_{3}}^{T}\left( Q(g)
\right) \sqrt{3}\frac{g_{A}}{N_{c}}\frac{k}{E_{\pi }}
\Biggr]
\,.
\label{A-calc-2}
\end{gather}

This expression is computed for the state (\ref{Bpi-TTJJ}), which is not an
eigenstate of $K$. Now we construct the eigenstates of $K$:
\begin{equation}
|TT_{3}JJ_{3}KB,\pi(k,L)\rangle=\sum\limits_{T_{B}}
\,|TT_{3}JJ_{3}B(T^{B}=J^{B}),\pi(k,L)\rangle\langle
T^{B}=J^{B}|K\rangle_{JTL}.
\label{K-TJ}
\end{equation}
The expression for the matrix elements $\langle T^{B}=J^{B}|K\rangle _{JTL}$
in terms of $6j$ symbols
can be read from Ref. \cite{HEHW-84}:
\begin{equation}
\langle T^{B}=J^{B}|K\rangle _{JTL}=(-1)^{L+T^{B}+J}\sqrt{\left(
2T^{B}+1\right) \left( 2K+1\right) }\left\{ 
\begin{array}{ccc}
K & T & J \\ 
T^{B} & L & 1
\end{array}
\right\} \,.  \label{ampl-K-TB-JB-via-6j}
\end{equation}
This leads to the expressions
\begin{align}
\langle T^{B}& =J^{B}|K=1\rangle _{JT,L=0}=(-1)^{T+1-J}\delta _{T^{B}J}\,,
\label{K-1-L-0-Mx}
\\
\langle T^{B}& =J^{B}|K=0\rangle _{JT,L=1}=\sqrt{\frac{2T^{B}+1}{3(2T+1)}}
\delta _{TJ}\,.
\label{K-0-L-1-Mx}
\end{align}
Now we derive from Eqs. (\ref{A-calc-2}), (\ref{K-TJ}), and (\ref{K-0-L-1-Mx}):
\begin{equation}
\Xi_{TT_{3}JJ_{3}B,\pi(k,L=1)}^{K=0}(g)=-i\sqrt{\pi}\frac{N_{c}^{1/2}}{F_{\pi}
}\delta_{TJ}D_{T_{3}J_{3}}^{T}\left[  Q(g)\right]  \frac{g_{A}}{N_{c}}
\frac{k}{E_{\pi}}\sum\limits_{T^{B}}\frac{2T^{B}+1}{\sqrt{2T+1}}\,.
\label{Xi-K0-L-1-calc}
\end{equation}
The summation over $T^{B}$ is restricted by the condition
\begin{equation}
|T-1|\leq T^{B}\leq T+1\,.
\end{equation}
At $T>1/2$ this leads to
\begin{equation}
\sum\limits_{T^{B}}\frac{2T^{B}+1}{\sqrt{2T+1}}
=\sum\limits_{T^{B}=T-1}^{T+1}\frac{2T^{B}+1}{\sqrt{2T+1}}
=3\sqrt{2T+1}\,.
\end{equation}
Actually we have 
\begin{equation}
\sum\limits_{T^{B}}\frac{2T^{B}+1}{\sqrt{2T+1}}=3\sqrt{2T+1}\,.
\label{sum-T-rad-T}
\end{equation}
for any $T$ (at $T=0$ the only possible value is $T^B=1$).

Inserting Eq. (\ref{sum-T-rad-T}) into Eq. (\ref{Xi-K0-L-1-calc}),
we obtain Eq.
(\ref{Xi-soft-K-0}). Similarly one derives Eq. (\ref{Xi-soft-K-1}) from Eqs.
(\ref{A-calc-2}), (\ref{K-TJ}), and (\ref{K-1-L-0-Mx}).

\end{document}